\RequirePackage{fix-cm}
\documentclass[smallextended]{svjour3}       
\smartqed  
\usepackage{graphicx}
\usepackage{natbib}
\usepackage{amsmath,amsfonts,amssymb,bm}
\usepackage{mathptmx}
\usepackage{newtxtext}
\usepackage{newtxmath}
\begin{document}

\title{Super-resolution data assimilation
}


\author{Sébastien Barthélémy        \and
        Julien Brajard \and
        Laurent Bertino \and
        François Counillon
}


\institute{Sébastien Barthélémy - Corresponding author \at
              Universitetet i Bergen, Bergen, Norway \\
              Bjerknes Center for Climate Research, Bergen, Norway \\
             \email{sebastien.barthelemy@uib.no}           
           \and
           Julien Brajard \at
              Nansen Environmental and Remote Sensing Center, Bergen, Norway \\
              Sorbonne Université, Paris, France \\
              \email{julien.brajard@nersc.no}
            \and
           Laurent Bertino \at
              Nansen Environmental and Remote Sensing Center, Bergen, Norway \\
              \email{laurent.bertino@nersc.no}
           \and
           François Counillon \at
              Nansen Environmental and Remote Sensing Center, Bergen, Norway \\
              Universitetet i Bergen, Bergen, Norway \\
              Bjerknes Center for Climate Research, Bergen, Norway \\
              \email{francois.counillon@nersc.no}
}

\date{Received: date / Accepted: date}

\maketitle

\begin{abstract}
Increasing the resolution of a model can improve the performance of a data assimilation system: first because model field are in better agreement with high resolution observations, then the corrections are better sustained and, with ensemble data assimilation, the forecast error covariances are improved. However, resolution increase is associated with a cubical increase of the computational costs. Here we are testing an approach inspired from images super-resolution techniques and called "Super-resolution data assimilation" (SRDA). Starting from a low-resolution forecast, a neural network (NN) emulates a high-resolution field that is then used to assimilate high-resolution observations. 
We apply the SRDA to a quasi-geostrophic model representing simplified surface ocean dynamics, with a model resolution up to four times lower than the reference high-resolution and we use the Ensemble Kalman Filter data assimilation method. We show that SRDA outperforms the low-resolution data assimilation approach and a SRDA version with cubic spline interpolation instead of NN. The NN's ability to anticipate the systematic differences between low and high resolution model dynamics explains the enhanced performance, for example by correcting the difference of propagation speed of eddies. Increasing the computational cost by 55\% above the LR data assimilation system (using a 25-members ensemble), the SRDA reduces the errors by 40\% making the performance very close to the HR system (16\% larger, compared to 92\% larger for the LR EnKF). The reliability of the ensemble system is not degraded by SRDA.
\end{abstract}

\section{Introduction}
\label{intro}
The quality of a forecast relates to the accuracy of the initial condition and its dynamical consistency. Data assimilation (DA) methods estimate such an initial condition based on observations, a dynamical model and statistical informations. The Ensemble Kalman Filter (EnKF, \citep{Evensen2003}) is one such method and consists of the recursive approach: a Monte Carlo model integration and a linear analysis update based on the ensemble covariance.

When designing a data assimilation system, computational resources are limited and there is a trade-off between assigning computing resources and increasing the cost of the data assimilation method ~\citep{lei2017}. On one hand, increasing the model resolution can (better) resolve small-scale processes that are parametrised otherwise, e.g. \citep{Gent1995}, and can as well enhances the skill of the larger scale \citep{Hewitt2017}. Increasing resolution can mitigate model bias (that are not well handled in data assimilation) and remove the need for representativity error \citep{Janjic2018} when assimilating high-resolution observations. On the other hand, increasing resolution can move the model into a gray zone of mixed parametrised/resolved resolution in particular for ocean models \citep{Hallberg2013} that are difficult to handle. Furthermore, there is an inverse cascade in the kinetic energy spectrum that lowers predictability for the small scale processes \citep{Sandery2017}. As such increasing the resolution requires a higher resolution observations network, larger ensemble size and higher frequency of assimilation to outperforms lower resolution systems \citep{Thoppil2021}. Improving the resolution also typically leads to a cubic computational cost which prohibits the use of more powerful data assimilation methods.

In practice, it is not uncommon for operational or climate centers to run concurrently two or more consecutive prototypes of the same ocean model at different resolutions as the old system keeps running during the developments of a new one.  However there are only few methods able to take systematic advantage from their coexistence: one coarse prototype that can afford an ensemble simulation, the other one that cannot. \citep{Gao2008} introduced a mixed-resolution data assimilation algorithm where the covariance is computed with a low-resolution dynamic ensemble, it is interpolated to the high-resolution grid and data assimilation is performed in the high-resolution space with only one member. The accuracy of the results depends on resolution discrepancies between the high and low-resolution models. \citep{Rainwater2013} introduced a mixed-resolution scheme based on the local ensemble transform Kalman filter, LETKF \citep{Hunt2007}, which makes use of two dynamic ensemble from a high and a low-resolution version of the Lorenz model and linearly combines their covariances matrix to update both ensembles. The authors have shown that for similar computational cost, the mixed-resolution scheme can achieve better results in terms of RMSE than the high-resolution standalone EnKF.

Another way to take advantage from a model with a lower resolution is super-resolution, that aims at increasing the resolution of an image. Super-resolution schemes based on a machine learning approach have been successfully applied in the field of the geosciences lately. For example, \citep{Rodrigues2018} used a convolutional neural network to provide high-resolution weather fields from low-resolution ones. The method was assessed over a region of South America and compared to a mean of a set of models at different resolutions, a linear regression between those models, and a regional model. The super-resolution strategy showed to improve the results in terms of RMSE compared to the three other methods. \citep{Vandal2018} provided a generalized stacked super resolution convolutional neural network framework for statistical downscaling of climate variables. It was tested with the super-resolution of precipitation fields over the contiguous United States and compared to other statistical downscaling methods. It was shown that this framework performed closely or better than the other selected methods, for example, in terms of daily predictability, extreme precipitations, daily root mean square error.

In this work, we propose an algorithm that consists in integrating the physical model in low resolution to produce the forecast and computing the analysis in high resolution. A neural network is used to map the low resolution forecast to a high-resolution field. If this mapping is accurate, our algorithm benefits from both a cheap model to integrate and a high-resolution analysis.

The overview of this article is as follows: the section~\ref{sect:method} presents the assimilation scheme used in this study and its combination with super-resolution. Section~\ref{sect:models_and_data} exposes the physical model used in this study, the neural network and the training set used, as well as the set-up of the data assimilation experiments. Sections~\ref{sect:results} displays the results while section~\ref{sect:discussion} discusses the results obtained. Section~\ref{sect:conclusion} provides the conclusions of this work together with some perspectives.

\section{Methods}
\label{sect:method}
\subsection{The deterministic ensemble Kalman filter}

Let $n$ be the model state dimension, $\bm{\mathsf{E}} \in \mathbb{R}^{n\times N}$ an ensemble of $N$ model states $\left(\mathbf{X}^{(1)},\mathbf{X}^{(2)},\dots,\mathbf{X}^{(N)} \right)$, $\mathbf{x} \in \mathbb{R}^{n}$ the ensemble mean and $\bm{\mathsf{A}} \in \mathbb{R}^{n\times N}$ the ensemble anomalies. $\mathbf{x}$ and $\bm{\mathsf{A}}$ are given by expressions~\eqref{eq:ens_mean} and~\eqref{eq:ens_anomalies} respectively.

\begin{subequations}
\begin{equation}
     \mathbf{x} = \frac{1}{N} \bm{\mathsf{E}} \bm{\mathsf{1}},
     \label{eq:ens_mean}
\end{equation}
\begin{equation}
     \bm{\mathsf{A}} = \bm{\mathsf{E}} \left(  \bm{\mathsf{I}} - \frac{1}{N}  \bm{\mathsf{1}}\bm{\mathsf{1}}^T\right),
     \label{eq:ens_anomalies}
\end{equation}
\end{subequations}

\parindent=0.cm where $\bm{\mathsf{I}} \in \mathbb{R}^{N\times N}$ is the identity matrix and $\bm{\mathsf{1}} \in \mathbb{R}^{N}$ is a vector with all elements equal to 1. In the following equations, the superscripts $\rm a$ and $\rm f$ stand respectively for the analysed and forecast states of the mean and the anomalies.

\parindent=0.3cm The true state of the system is noted $\mathbf{x}^{\rm t}$. In this study, we assimilate synthetic observations $\mathbf{d}$ that are defined by: 

\begin{equation*}
    \mathbf{d} = \bm{\mathsf{H}} \mathbf{x}^{\rm t}+ \boldsymbol\varepsilon, \quad \boldsymbol\varepsilon \sim \mathcal{N}\left( \mathbf{0}, \bm{\mathsf{R}}\right)
    \label{eq:observations}
\end{equation*}

\parindent=0cm where $\bm{\mathsf{H}} \in \mathbb{R}^{p\times n}$ ($p$ being the number of observations) is such that for each grid point $j$, $\bm{\mathsf{H}}_{ij} = 1$ if $i$ corresponds to the observation point, 0 otherwise. $\bm{\mathsf{R}}$ is the observation error covariance matrix. The observation errors are uncorrelated and have for standard deviation the same value $\sigma_o$. Thus, $\bm{\mathsf{R}}$ is diagonal with $\sigma_o^2$ on the diagonal. 

\parindent=0.3cm In this study, the deterministic EnKF (DEnKF) introduced by \citet{Sakov2008} is used. The DEnKF is a square-root (deterministic) formulation of the EnKF that solves the analysis without the need for perturbation of the observations. It inflates the errors by construction and is intended to perform well in operational applications where corrections are small \citep{Sakov2008}. The DEnKF uses the two steps of sequential data assimilation: a forecast step and an analysis step.

In the forecast step, each member $i$ is integrated by the model from one assimilation cycle $k-1$ to the next $k$:

\begin{equation}
    \mathbf{X}^{{\rm f},i}_k = \mathcal{M} \left(\mathbf{X}^{{\rm a},i}_{k-1} \right), \quad i = 1,\dots,N,
    \label{eq:model_integration}
\end{equation}

\parindent=0.cm where $\mathcal{M}$ is an operator that stands for the model integration.

\parindent=0.3cm The analysis step of the DEnKF proceeds at assimilation cycle $k$ in two stages, the update of the mean, Eq.~(\ref{eq:update_mean}), and the update of the ensemble anomalies, Eq.~(\ref{eq:update_anomalies}):

\begin{subequations}
\begin{equation}
     \mathbf{x}^{\rm a}_k = \mathbf{x}^{\rm f}_k + \bm{\mathsf{K}}_k \left( \mathbf{d}_k - \bm{\mathsf{H}}_k \mathbf{x}^{\rm f}_k \right),
     \label{eq:update_mean}
\end{equation}
\begin{equation}
     \bm{\mathsf{A}}^{\rm a}_k = \bm{\mathsf{A}}^{\rm f}_k -\frac{1}{2} \bm{\mathsf{K}}_k \bm{\mathsf{H}}_k \bm{\mathsf{A}}^{\rm f}_k,
     \label{eq:update_anomalies}
\end{equation}
\end{subequations}

\parindent=0.cm where :

\begin{subequations}
\begin{equation}
    \bm{\mathsf{K}}_k = \bm{\mathsf{P}}^{\rm f}_k \bm{\mathsf{H}}_k^T \left( \bm{\mathsf{H}}_k \bm{\mathsf{P}}^{\rm f}_k \bm{\mathsf{H}}_k^T + \bm{\mathsf{R}}\right)^{-1},
    \label{eq:kalman_gain}
\end{equation}
\begin{equation}
    \bm{\mathsf{P}}^{\rm f}_k = \frac{\bm{\mathsf{A}}_k^{\rm f} \left( \bm{\mathsf{A}}_k^{\rm f}\right)^T}{N-1}
\end{equation}
\end{subequations}

are respectively the Kalman gain matrix and the background error covariance matrix estimated from the ensemble anomalies.

\parindent=0.3cm 

\parindent=0.3cm The DEnKF method will be used in the following but will be referred to as EnKF since the algorithmic parts of SRDA are independent of the flavour of the EnKF analysis scheme.

\subsection{Super-resolution data assimilation}\label{sect:SRDA}

The super-resolution method involves a model at two different resolutions, high (HR) and low (LR). In the following, the subscripts $H$ and $L$ will be used in the equations to denote if an object, matrix or vector, is in the HR or the LR space. The principle of super-resolution is to perform the forecast step with the LR model in order to reduce the computational cost of that step and to perform the analysis step in the HR space to benefit from HR observations rather than LR observations to update the ensemble. 

Let $\bm{\mathsf{E}}_L$ be an ensemble of $N$ LR model states $\left(\mathbf{X}_{L}^{(1)},\mathbf{X}_{L}^{(2)},\dots,\mathbf{X}_{L}^{(N)} \right)$. Following Eq.~\eqref{eq:model_integration}, the forecast step from assimilation cycle $k-1$ to assimilation $k$ writes for each ensemble member:

\begin{equation}
    \mathbf{X}^{{\rm f},i}_{L,k} = \mathcal{M}_L \left(\mathbf{X}^{{\rm a},i}_{L,k-1} \right), \quad i = 1,\dots,N,
    \label{eq:model_integration_LR}
\end{equation}
 
\parindent=0cm where $\mathcal{M}_L$ stands for the LR model integration. 

\parindent=0.3cm At the end of the forecast step every member $\mathbf{X}^{{\rm f},i}_{L,k}$ is downscaled from the LR to the HR grid to produce an emulated HR member $\mathbf{X}^{{\rm f},i}_{H,k}$ :

\begin{equation}
    \mathbf{X}^{{\rm f},i}_{H,k} = \mathcal{D} \left(\mathbf{X}^{{\rm f},i}_{L,k}\right), \quad i = 1,\dots,N,
    \label{eq:downscaling}
\end{equation}

\parindent=0cm where $\mathcal{D}$ stands for a downscaling operator from the LR to the HR grid. In this study, two different downscaling operators are used: a cubic spline interpolation operator and a neural network (hereafter denoted NN).

\parindent=0.3cm The mean $\mathbf{x}_{H,k}$ and the anomalies $\bm{\mathsf{A}}_{H,k}$ of the emulated HR ensemble at assimilation cycle $k$ are updated based on Eq.~(\ref{eq:update_mean}) and (\ref{eq:update_anomalies}):

\begin{subequations}
\begin{equation}
     \mathbf{x}^{\rm a}_{H,k} = \mathbf{x}^{\rm f}_{H,k} + \bm{\mathsf{K}}_{H,k} \left( \mathbf{d}_{H,k} - \bm{\mathsf{H}}_{H,k} \mathbf{x}^{\rm f}_{H,k} \right),
     \label{eq:update_mean_HR}
\end{equation}
\begin{equation}
     \bm{\mathsf{A}}^{\rm a}_{H,k} = \bm{\mathsf{A}}^{\rm f}_{H,k} -\frac{1}{2} \bm{\mathsf{K}}_{H,k} \bm{\mathsf{H}}_{H,k} \bm{\mathsf{A}}^{\rm f}_{H,k},
     \label{eq:update_anomalies_HR}
\end{equation}
\end{subequations}

\parindent=0.cm where $\bm{\mathsf{K}}_{H,k}$, $\bm{\mathsf{H}}_{H,k}$ and $\mathbf{d}_{H,k}$ are respectively the HR Kalman gain, the HR observation operator and the HR observations at assimilation cycle $k$. 

\parindent=0.3cm After the analysis step, every member of the HR ensemble is upscaled from the HR grid back to the LR grid before the following forecast step:

\begin{equation}
    \mathbf{X}^{{\rm a},i}_{L,k} = \mathcal{U} \left(\mathbf{X}^{{\rm a},i}_{H,k}\right), \quad i = 1,\dots,N,
    \label{eq:upscaling}
\end{equation}

\parindent=0.cm where $\mathcal{U}$ is the upscaling operator from the HR to the LR grid. In this study, $\mathcal{U}$ is always a cubic spline interpolation operator.

\parindent=0.3cm In this study, we have used also an ultra-low resolution version of the model, referred to as ULR (see section \ref{sect:models_and_data}-\ref{sect:qg_model}), instead of the LR version. In this case, equations (\ref{eq:model_integration_LR}), (\ref{eq:upscaling}) and (\ref{eq:downscaling}) still apply the same.

In the following, depending on the choice of the downscaling method the super-resolution data assimilation scheme will be referred to as SRDA-NN or SRDA-cubic, or simply SRDA if there is no need to specify the downscaling method.

\subsection{Rewriting the SRDA as a LR scheme}\label{sect:LR-SRDA}

In the particular case where the LR and HR grids overlap, the upscaling operator $\mathcal{U}$ is equivalent to a sub-sampling operator and is thus linear. It can be showed that the upscaling of the Eq.~(\ref{eq:update_mean_HR})-(\ref{eq:update_anomalies_HR}) to the LR grid with the operator $\mathcal{U}$ leads to the following system of LR equations (see Appendix~\ref{app:lr_srda} for details):


\begin{subequations}
\begin{equation}
    \mathbf{x}_{L,k}^{\rm a} = \tilde{\mathbf{x}}_{L,k}^{\rm f} + \frac{\tilde{\mathbf{A}}_{L,k}^{\rm f} \hat{\mathbf{A}}_{H,k}^{\rm f}}{N-1} \left( \frac{\hat{\mathbf{A}}_{H,k}^{\rm f} \left(\hat{\mathbf{A}}_{H,k}^{\rm f}\right)^T}{N-1} + \mathbf{R} \right)^{-1} \left( \mathbf{d}_{H,k} - \hat{\mathbf{x}}_{H,k}^{\rm f}\right), \label{eq:srda_lr_mean} 
\end{equation}
\begin{equation}
    \mathbf{A}_{L,k}^{\rm a} = \tilde{\mathbf{A}}_{L,k}^{\rm f} - \frac{1}{2} \frac{\tilde{\mathbf{A}}_{L,k}^{\rm f}  \hat{\mathbf{A}}_{H,k}^{\rm f}}{N-1} \left( \frac{\hat{\mathbf{A}}_{H,k}^{\rm f} \left(\hat{\mathbf{A}}_{H,k}^{\rm f}\right)^T}{N-1} + \mathbf{R} \right)^{-1} \hat{\mathbf{A}}_{H,k}^{\rm f}, \label{eq:srda_lr_anom}
\end{equation}
\end{subequations}

\parindent=0cm where $\tilde{\mathbf{x}}_{L,k}^{\rm f}$ and $\tilde{\mathbf{A}}_{L,k}^{\rm f}$ are respectively the mean and the anomalies of a LR ensemble obtained after application of an operator $\mathcal{Q}$ on the LR background ensemble $\mathbf{E}_{L,k}^{\rm f}$; $\mathcal{Q}$ stands for the correction of the LR model error (see Appendix~\ref{app:lr_srda}). We have:

\begin{equation*}
\tilde{\mathbf{x}}_{L,k}^{\rm f} = \overline{\mathcal{Q}\left( \mathbf{E}_{L,k}^{\rm f}\right)}, \quad \tilde{\mathbf{A}}_{L,k}^{\rm f} = \mathcal{Q}\left( \mathbf{E}_{L,k}^{\rm f}\right) - \overline{\mathcal{Q}\left( \mathbf{E}_{L,k}^{\rm f}\right)}.
\end{equation*}

\parindent=0.3cm  Similarly, $\hat{\mathbf{x}}_{H,k}^{\rm f}$ and $\hat{\mathbf{A}}_{H,k}^{\rm f}$ are respectively the mean and the anomalies of the LR ensemble mapped to the HR observations' space after application of a super-resolution observation operator $\mathcal{H}$ on the LR background ensemble $\mathbf{E}_{L,k}^{\rm f}$ (see Appendix~\ref{app:lr_srda}). We have:

\begin{equation*}
\hat{\mathbf{x}}_{H,k}^{\rm f} = \overline{\mathcal{H}\left( \mathbf{E}_{L,k}^{\rm f}\right)}, \quad  \hat{\mathbf{A}}_{H,k}^{\rm f} = \mathcal{H}\left( \mathbf{E}_{L,k}^{\rm f}\right) - \overline{\mathcal{H}\left( \mathbf{E}_{L,k}^{\rm f}\right)}.
\end{equation*}

Following Eq.~(\ref{eq:srda_lr_mean})-(\ref{eq:srda_lr_anom}), the SRDA can be interpreted as the combination of the correction of the LR model error and the assimilation of HR observations with a super-resolution observation operator. On one hand, this formulation of the SRDA has the advantage that it does not require to downscale and upscale the whole ensemble at each assimilation cycle, it only requires the downscaling of the observed fields at the observation points. But on the other hand, it requires two different operators, the operator $\mathcal{Q}$ to correct the LR field and the super-resolution operator $\mathcal{L}$. Additionally, the analysis is in LR instead of HR. In the following, the HR formulation of the SRDA is used.

\section{Models and data}\label{sect:models_and_data}

\subsection{Quasi-geostrophic model}\label{sect:qg_model}

\parindent=0.3cm The version of the QG model used here follows the set up described in \citep{Sakov2008}. It is a 1.5-layer reduced-gravity QG model with double-gyre wind forcing and biharmonic friction. For more details about the QG model, see \citet{Sakov2008}. In this study, the QG model is used at three different resolution that are summarized in Table~\ref{tab:model-config}. In the table, the grid point size is expressed in terms of HR grid points, and the time step is expressed in terms of that of the HR model. The domain used in this study is $129\times129$. The computational cost of doubling the resolution from LR to HR results in increasing the computational cost by a factor 8 as there are 4 times more points (in x and y) and one needs to divide the time step by two in order to satisfy the Courant-Friedrichs-Lewy condition. An example of output of the QG model is given in figure~\ref{fig:model_snapshot}.

\begin{table}[ht]
\caption{Summary of different model configurations}\label{tab:model-config}
\begin{center}
\begin{tabular}{cccccc}
\hline\noalign{\smallskip}
Name & Grid point size & Time step & State size & Snapshot figure\\
\noalign{\smallskip}\hline\noalign{\smallskip}
 HR & 1 & 1 & $129\times 129$ & Fig.~\ref{fig:model_snapshot}-(a)\\
 LR & 4 & 2 & $65\times 65$ & Fig.~\ref{fig:model_snapshot}-(b)\\
 ULR & 16 & 4 & $33\times 33$ & Fig.~\ref{fig:model_snapshot}-(c)\\
\end{tabular}
\end{center}
\end{table}

\subsection{Training data set}
The neural network (NN) super-resolution operator $\mathcal{D}_{HL}$ is trained using a dataset obtained from a 120,000 time step simulation of the HR model.
From this simulation, $K$ ($K=10,001$ in our case) snapshots $\mathbf{X}^{\rm f}_{H,k}$ were regularly sampled at time steps $t_k$, such as $t_k-t_{k-1} =  \Delta t$ ($\Delta t = 12$ in our case). At each time step $t_{k-1}$, the HR-field is first scaled to low-dimension using the operator $\mathcal{U}$ as defined in Eq.~(\ref{eq:upscaling}), and then the LR model is integrated over $\Delta t$, to produce a low-resolution field $\mathbf{X}^{\rm f}_{L,k}$ such as
\begin{equation}
   \mathbf{X}_{L,k} = \mathcal{M}_L \circ \mathcal{U}\left(\mathbf{X}_{H,(k-1)} \right).
    \label{eq:training-lr}
\end{equation}
The 10,000 couples $\left(\mathbf{X}_{L,k}, \mathbf{X}_{H,k}\right)$ can be used to calibrate a NN super-resolution operator. The first 8,000 samples of the dataset constitute the training set, used to optimize the parameters of the neural network (see section~\ref{sect:nn}). The last 2,000 samples are the validation set, used to evaluate the performances of the operator.
We applied the same procedure to produce the dataset in the ULR case.

Note that, while this procedure aims at mimicking the low-resolution forecast members $\mathbf{X}^{\rm f}_{L,k}$ obtained in the SRDA approach, the statistical distribution of the forecast could be slightly different in the training phase as during the data assimilation. Indeed in the training/validation, the LR forecast is initialized with an upscaled HR simulation, whereas during the SRDA algorithm the forecast is initialized with an upscaled analysis. If the model contains bias, the analysis states can be statistically different from the states obtained by a HR model. In this is the case, an additional DA step could be performed to produce the training set, similarly to what is done in~\citet{brajard2021combining}. 

\subsection{Super-resolution neural network}
\label{sect:nn}
\begin{figure}[ht]
 \begin{center}
    \includegraphics[width=\textwidth]{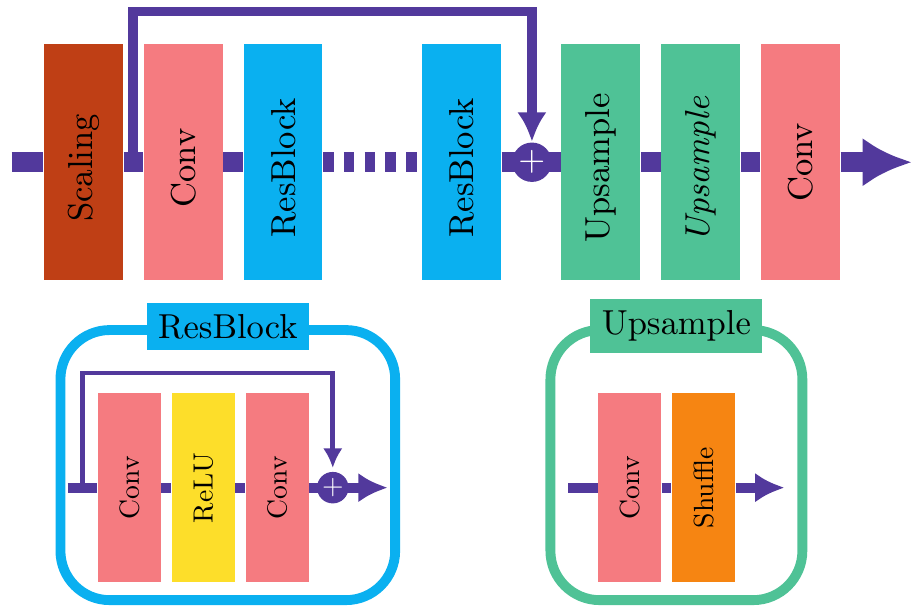}
 \end{center}
 \caption{Diagram representing the NN model. The italic {\it Upsample} block is used only for upsample from ULR to HR, otherwise only one Upsample block is used.}
 \label{fig:edsr}
\end{figure}

The neural network architecture is the enhanced deep super-resolution network (EDSR) adapted from~\cite{Lim_2017_CVPR_Workshops} in which super-resolution is applied to RGB photographic images. However, geophysical fields present differences with photographic images. First, the number of channels is not fixed to 3 (Red, Green, Blue). In the case of the quasi-geostrophic model, there is only one channel corresponding to the sea level elevation. Second, the pattern and texture of sea level elevation are smoother and more homogeneous than patterns in photographic images. For those reasons, the network chosen has been slightly simplified by specifying only one input channel and by reducing the number of degrees of freedoms (hereafter denoted {\it weights}) in comparison with~\cite{Lim_2017_CVPR_Workshops}.
A diagram of the network is presented in figure~\ref{fig:edsr}. We give hereafter a short description of all the blocks, a more complete description can be found in~\citet{Lim_2017_CVPR_Workshops}.

The {\bf Scaling} block is a simple scaling function that scales the input features images. In our case, it is a multiplicative factor of 0.04 so that the values are mainly between -1 and 1. There are no trainable weights in this block.

The {\bf Conv} block is a convolutional block as introduced in~\cite{lecun1989generalization}. In our model, each convolutional block is composed of 16 filters of size $3\times3$. The weights of the filters are optimized during the training phase.

The {\bf ReLU} block is a a nonlinear function $f$ defined by $f(x)=\max(0,x)$ applied point-wise to each input feature. It enables the model to be non-linear.

The {\bf ResBlock} is a non-linear convolutive block in which the input is added to the output of the block. In particular, it mitigates the gradient vanishing problem in deep learning architecture and it has been proved to be efficient in image processing problems~\citep{he2016deep}. The weights to be optimized are the weights of the filter of the convolutional blocks.

The {\bf Shuffle} is where the scaling is realized in practice. The block takes as inputs $F\times2^2$ image features of size $n_L\times n_L$ where $n_L$ is the size of the low-resolution image ($64\times64$ for LR and $32\times32$ for ULR) and produce $F$ image feature of size $n_H\times n_H$ where $n_H=2\times n_L$ by intertwines input features into spatial blocks. There are no trainable weights in this block. This procedure is detailed in~\cite{shi2016real}.

The total number of weights of the neural network is 22,273 for the model mapping LR to HR, and 23,361 for the model mapping ULR to HR.

\begin{figure}[ht]
 \begin{center}
    \includegraphics[width=\textwidth]{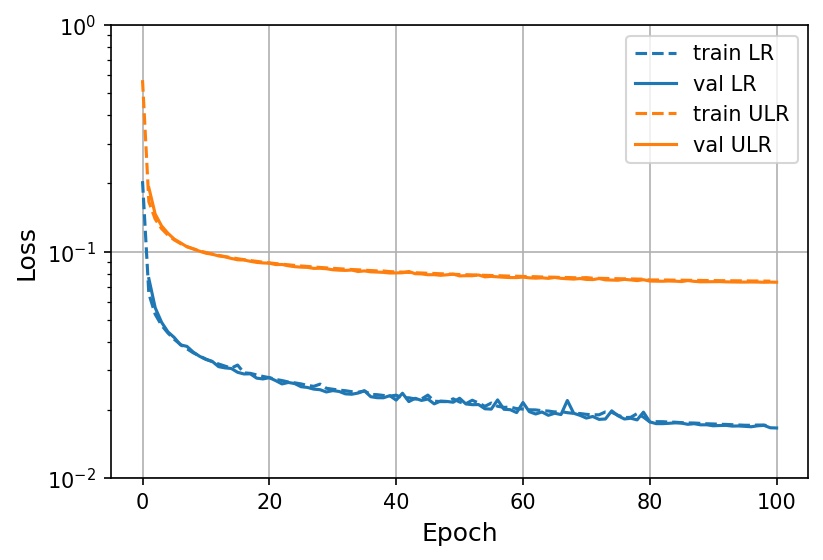}
 \end{center}
 \caption{Convergence of the training of the NN models. Solid lines represents the loss of the validation set and the dashed line the loss of the training set. The blue line is the loss of the LR model and the orange line is the loss of the ULR model.}
 \label{fig:learning-curve}
\end{figure}

The training of the model consists of optimizing the weights of the convolutional blocks to minimize the mean absolute error between the output and the target contained in the training set. The optimizer chosen is Adam~\citep{kingma2014adam}, with a learning rate of $10^{-4}$ and a batch size of 32. The training is stopped after 100 epochs (an epoch is when all the training samples have been presented to the model). Figure~\ref{fig:learning-curve} shows the evolution of the loss (mean absolute error) as a function of the epochs. It can be seen that the decreasing of the loss has been stabilized after 100 epochs for both models (ULR and LR). Very little tuning of the training was performed because most of the settings were chosen from~\cite{Lim_2017_CVPR_Workshops}. Fine-tuning would be possible and could help to reduce the cost of the training and of the model computation, but it was found to be unnecessary here given the relative simplicity of the physical system considered.

\subsection{Set-up of the data assimilation experiments}

We have use a twin experiment to assess the validity of the SRDA scheme. The experiment is carried over a time window of 6000 time steps with an analysis step every 12 time steps, which amounts to a total of 500 assimilation cycles. 12 time steps corresponds also to the frequency of the model outputs in the training of the neural network. 

In order to evaluate the benefits retrieved from the SRDA scheme, it was compared to the usual EnKF on the LR and the ULR grids (referred to hereafter as EnKF-LR and EnKF-ULR respectively). It was also compared to the usual EnKF on the HR grid (referred to as EnKF-HR).

The true run is generated with the HR model and with a viscosity coefficient $\nu_t = 2 \times 10^{-12}$ while the viscosity, $\nu$, for the data assimilation experiments is set to $\nu = 2 \times 10^{-11}$ for the 3 different resolutions of the model. We used two different values of the viscosity because we followed the framework of \citep{Sakov2008} where the authors have used those values of the viscosity in order to "achieve stable performances".

\begin{figure}[ht]
 \begin{center}
     \begin{tabular}{c}
    \includegraphics[width=19pc]{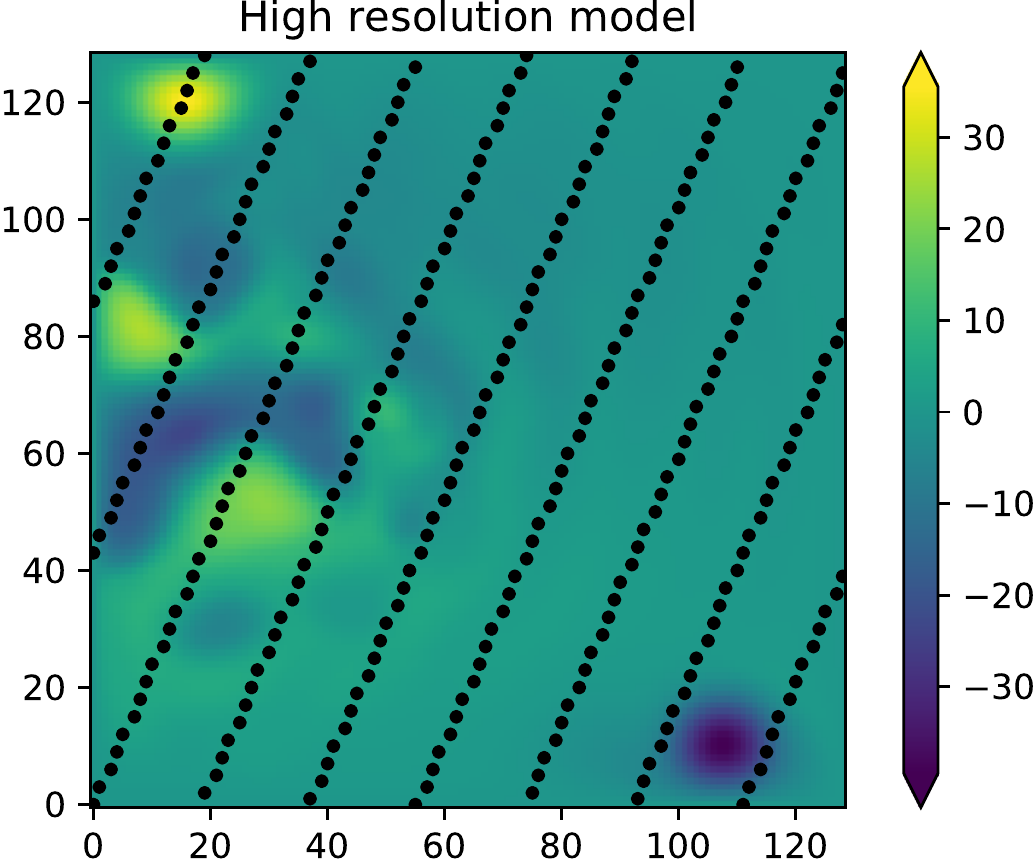} \\
    (a) \\
    \includegraphics[width=19pc]{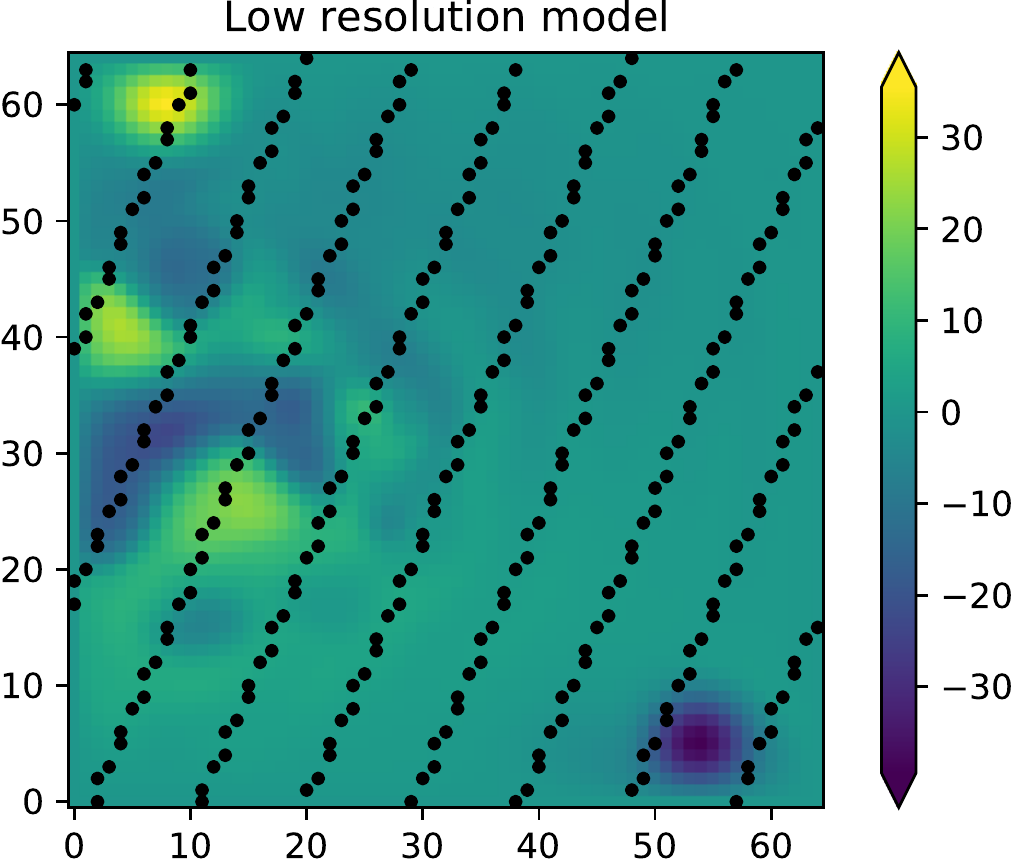} \\
    (b) \\
    \includegraphics[width=19pc]{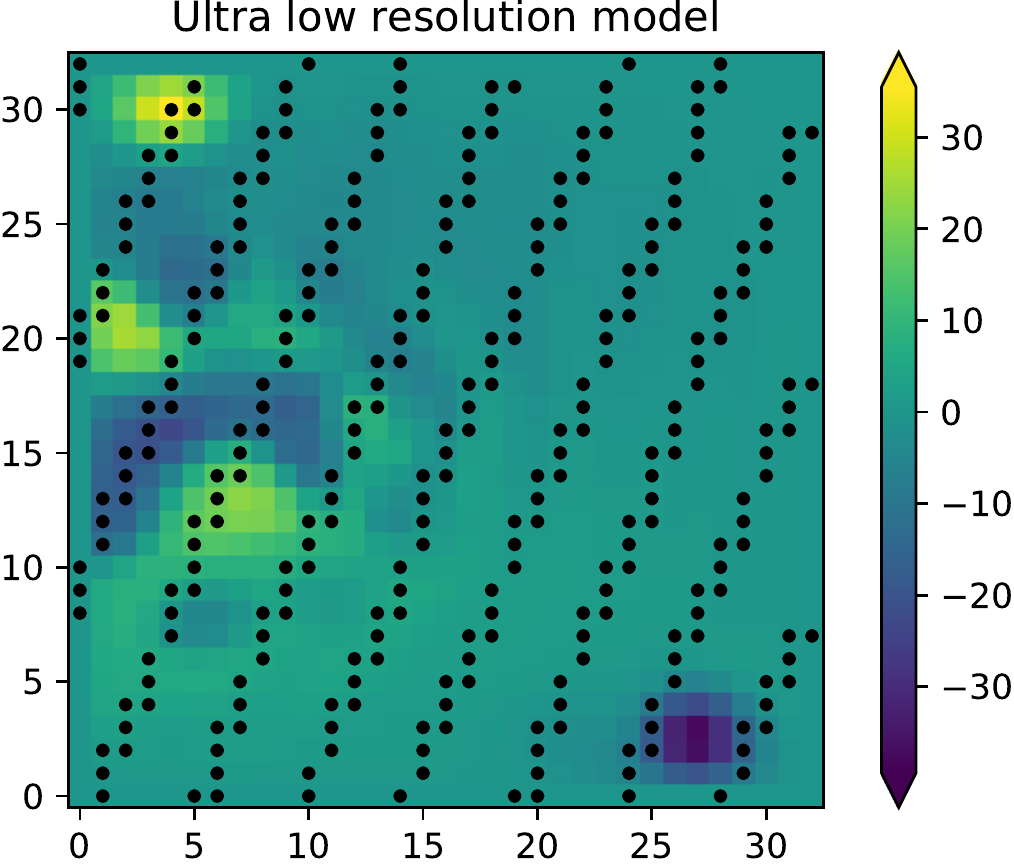} \\
    (c) 
 \end{tabular}
 \end{center}
 \caption{Snapshot of the sea level elevation for (a) the HR model (b) the LR model and (c) the ULR model. The black points stand for the location of the observation points at a given time}
 \label{fig:model_snapshot}
\end{figure}

The observations are generated by adding to the true run a Gaussian noise with 0 mean and standard-deviation $\sigma_{o,H} = 2$. There are 300 observation points. The location of the observation points mimics a satellite track and slightly differs from one assimilation cycle to the other, see figure~\ref{fig:model_snapshot}-(a) for illustration. The LR (resp. ULR) observations are the same as those of the true run and their location is derived from that of the HR grid by shifting each observation point to the nearest LR (resp. ULR) grid point. In the case where two observations are shifted to the same grid point, the point with the highest initial ordinate is shifted to the point above (see figures~\ref{fig:model_snapshot}-(b) and (c)) in order to avoid two observations to overlap. This results in an increase of the observation error over the LR and the ULR grids, respectively: $\sigma_{o,L} = 2.4$ and $\sigma_{o,UL} = 3.7$. In order to mitigate the impact of sampling errors on the results, a local analysis scheme using a Gaspari and Cohn function as a tapering function, was used~\citep{Sakov2011}. This scheme is equivalent to artificially increasing the observation error of distant observations in order to take into account only local observations when updating the ensemble at a given point.

At each assimilation cycle $k$, the performance of the different assimilation schemes were estimated using the spatial root mean square error $r_k$, Eq.~(\ref{eq:rmse}), the ensemble spread $s_k$ \citep{Fortin2014}, Eq.~(\ref{eq:sqrt_mean_var}), and the Pearson correlation coefficient $c_k$, Eq.~(\ref{eq:correlation}).

\begin{equation}
    r_k = \sqrt{\frac{1}{n}\sum_{j=1}^n \left( \mathbf{x}^{\rm a}_k(j) - \mathbf{x}^{\rm t}_k(j)\right)^2},
    \label{eq:rmse}
\end{equation}

\begin{equation}
    s_k = \sqrt{\frac{1}{n} \sum_{j=1}^n \frac{1}{N-1}\sum_{i=1}^{N} \left( \bm{\mathsf{X}}_k^{{\rm a},i}(j)-\mathbf{x}_k^{\rm a}(j)\right)^2},
    \label{eq:sqrt_mean_var}
\end{equation}

\begin{equation}
    c_k = \frac{\sum_{j=1}^n \left( \mathbf{x}_k^{\rm a}(j) - \bar{\mathbf{x}}_k^{\rm a}\right) \left( \mathbf{x}_k^{\rm t}(j) - \bar{\mathbf{x}}_k^{\rm t} \right)}{\sqrt{\sum_{j=1}^n \left( \mathbf{x}_k^{\rm a}(j) - \bar{\mathbf{x}}_k^{\rm a}\right)^2} \sqrt{\sum_{j=1}^n \left( \mathbf{x}_k^{\rm t}(j) - \bar{\mathbf{x}}_k^{\rm t} \right)^2}},
    \label{eq:correlation}
\end{equation}

\parindent=0cm where $\bar{\mathbf{x}}_k^{\rm a} = \frac{1}{n} \sum_{j=1}^n \mathbf{x}_k^{\rm a}(j)$ and $\bar{\mathbf{x}}_k^{\rm t} = \frac{1}{n} \sum_{j=1}^n \mathbf{x}_k^{\rm t}(j)$.

\parindent=0.3cm The temporal mean of these scores, computed over the assimilation cycles after cycle 10, are also considered. The first 10 assimilation cycles correspond to a spin-up period and are not relevant for the computation of the mean of these scores.

As with the EnKF-HR, the SRDA-NN and the SRDA-cubic schemes, the assimilation step is performed in the HR space, the scores $r$, $s$ and $c$, are computed in the HR space while for the EnKF-LR scheme (EnKF-ULR respectively), the scores are computed in the LR (ULR) space. As these scores are averaged over the size of the domain ($r$ and $s$) or do not depend on it ($c$), it allows for a comparison of the scores despite the difference in resolution of the different schemes.

\section{Results} \label{sect:results}

\subsection{Super-resolution} \label{sect:super-resolution}
\begin{figure}[ht]
 \begin{center}
    \includegraphics[width=\textwidth]{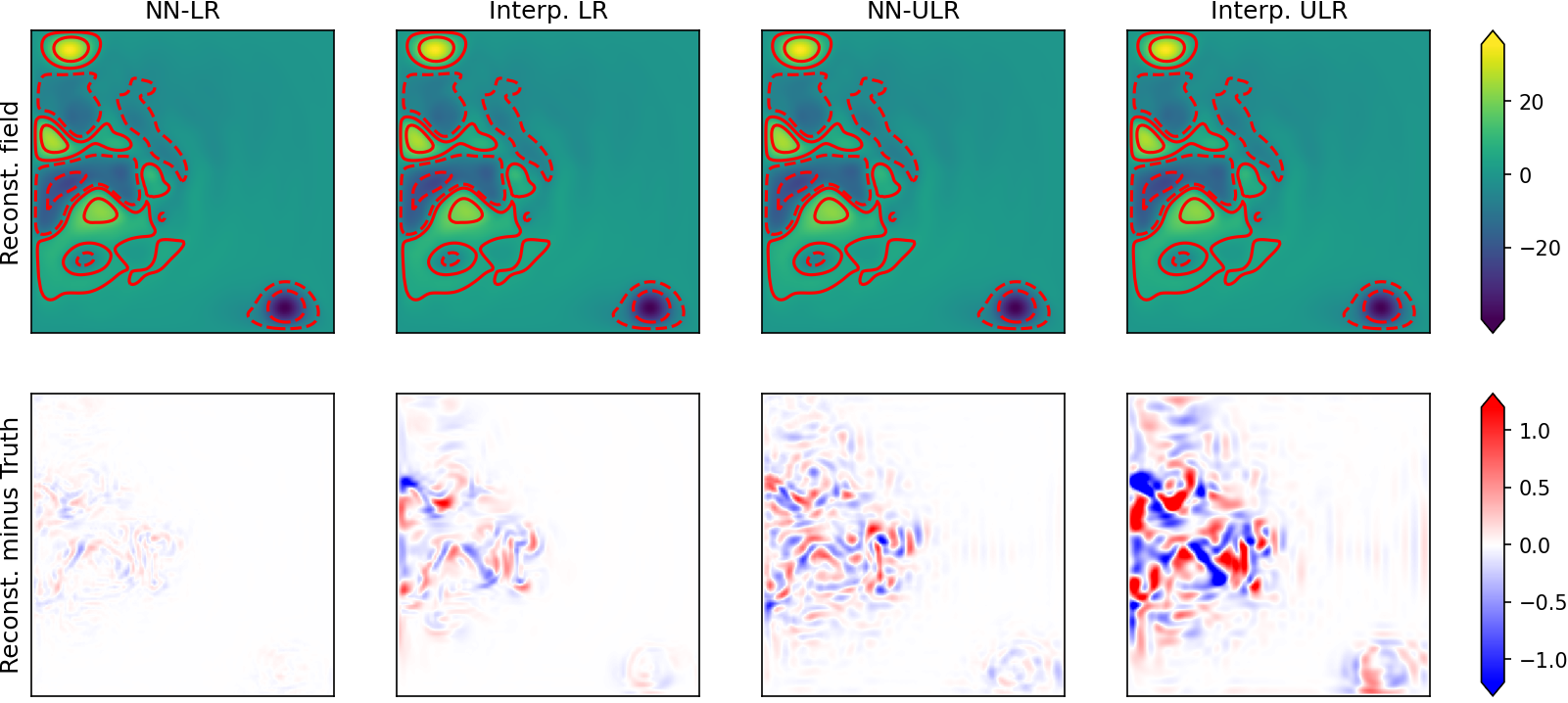}
 \end{center}
 \caption{Snapshot (top row) of the HR-reconstructed sea-level elevation from a LR field and ULR field computed by the cubic spline interpolation and by the neural network. The contour (with values of -18, -4 in dotted lines and 4, 18 in solid lines) correspond to the true state.  The bottom row represents the difference between the reconstructed field and the true reference field. The snapshot belongs to the validation set (not used for optimizing the NN).}
 \label{fig:compare_sample}
\end{figure}

 Examples of HR fields reconstructed from LR and ULR are shown in figure~\ref{fig:compare_sample}. The original high-resolution field has been arbitrarily chosen in the validation dataset. The reconstructions by a cubic-spline interpolation and the NN model are compared. Due to the spatial variability, the reconstructed fields are not easily distinguishable and  appears very alike to the true HR field (represented in contour plot). Nevertheless, looking at the difference between the reconstructed field and the truth (second raw in Fig.~\ref{fig:compare_sample}), we can see that NN reduces significantly the error of the interpolation. As expected, the error of the reconstruction is higher from ULR than from LR and the error is more important in dynamical active regions. Another remarkable feature can be noticed by looking at the eddy situated South-East of the field. The bimodal error pattern visible in the interpolation reconstruction is typical of a displacement error. The eddy in the interpolation reconstructed field is located to the east of the true location. This demonstrates that low-resolution models have a biases in the eddy motion speed. The NN model corrects most of this error, showing that, in addition to reducing error in the small scale features, it can also mitigate some systematic biases.

\begin{figure}[ht]
 \begin{center}
    \includegraphics[width=.5\textwidth]{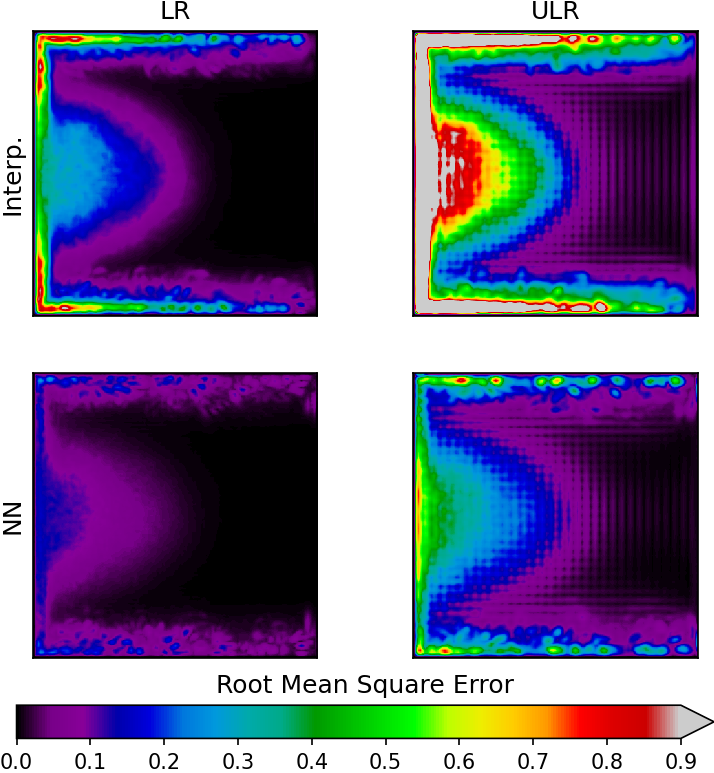}
 \end{center}
 \caption{Root-mean square error of the HR-reconstructed sea-level elevation from the LR field (left column) and from the ULR field (right column) by the cubic spline interpolation (top row) and the neural network (bottom row). The error is computed on the validation set (not used for optimisation).}
 \label{fig:compare_rmse}
\end{figure}

In figure~\ref{fig:compare_rmse}, the root-mean-squared error averaged over the whole validation period is presented. It confirms that there is a significant overall improvement in the reconstruction with NN. This result also confirms that the improvement is most noticeable in the high mesoscale activity region.

\subsection{Data assimilation}

The data assimilation experiments are performed with the different assimilation schemes (EnKF-HR, EnKF-LR, SRDA-NN, SRDA-cubic), and at different  ensemble size. For each experiment, a sensitivity analysis was performed in identify the optimal inflation coefficient and localization radius (in terms of RMSE), which allows a fair comparison of the different schemes. While the optimal inflation coefficient is roughly the same for all the schemes (results not showed), the optimal localization radius of the EnKF-HR is larger than that of all the  other schemes. It shows in particular that the SRDA-NN is prompt to larger spurious covariances than the EnKF-HR. In the following, the results exhibited are those obtained with the optimal parameters unless explicitly stated. This general sensitivity analysis was performed for both a LR ensemble forecast and an ULR ensemble forecast. 

\begin{figure}[ht]
\hspace{-0.5cm}
    \begin{tabular}{cc}
    \includegraphics[width=0.5\textwidth]{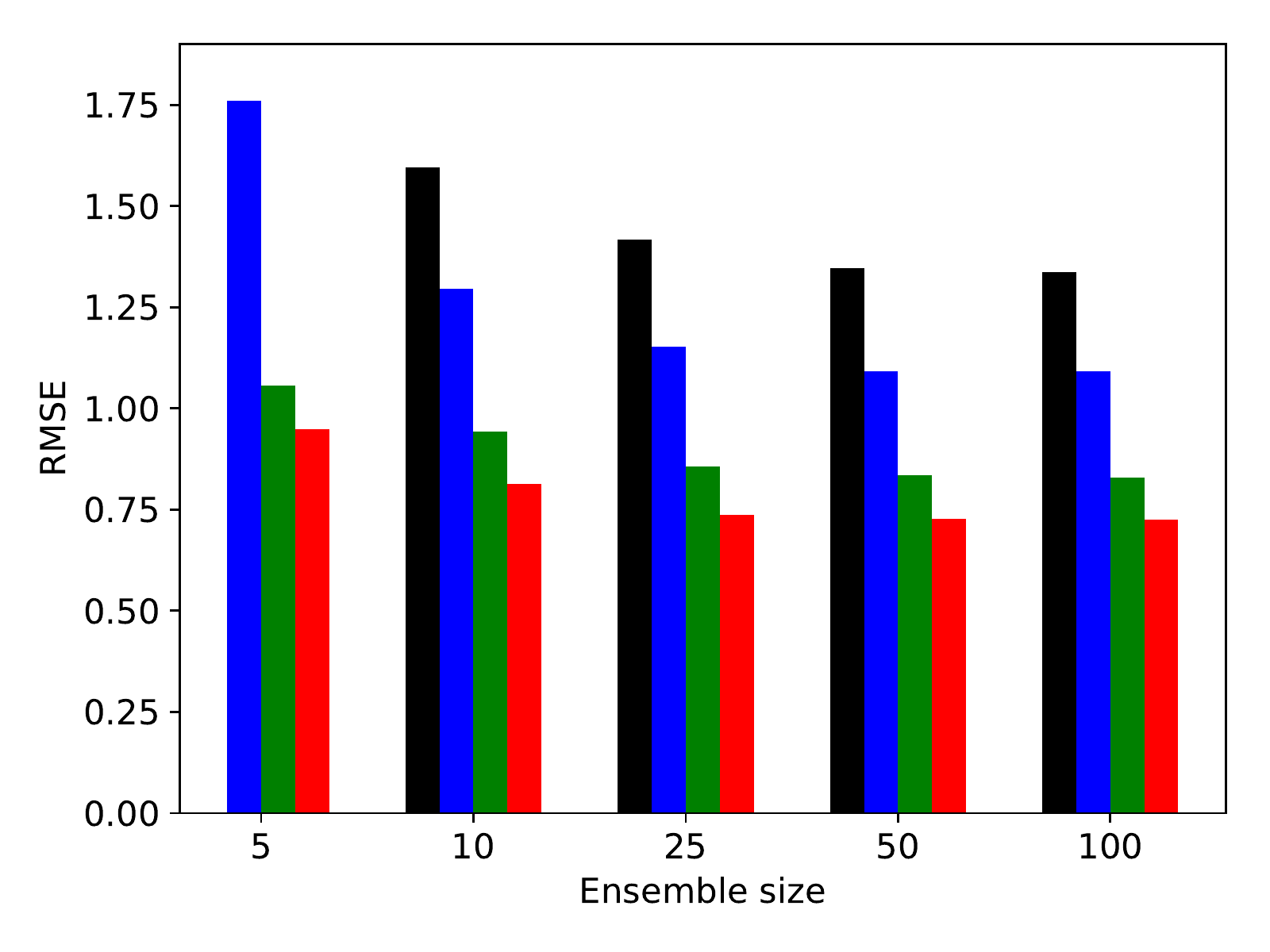} & \includegraphics[width=0.5\textwidth]{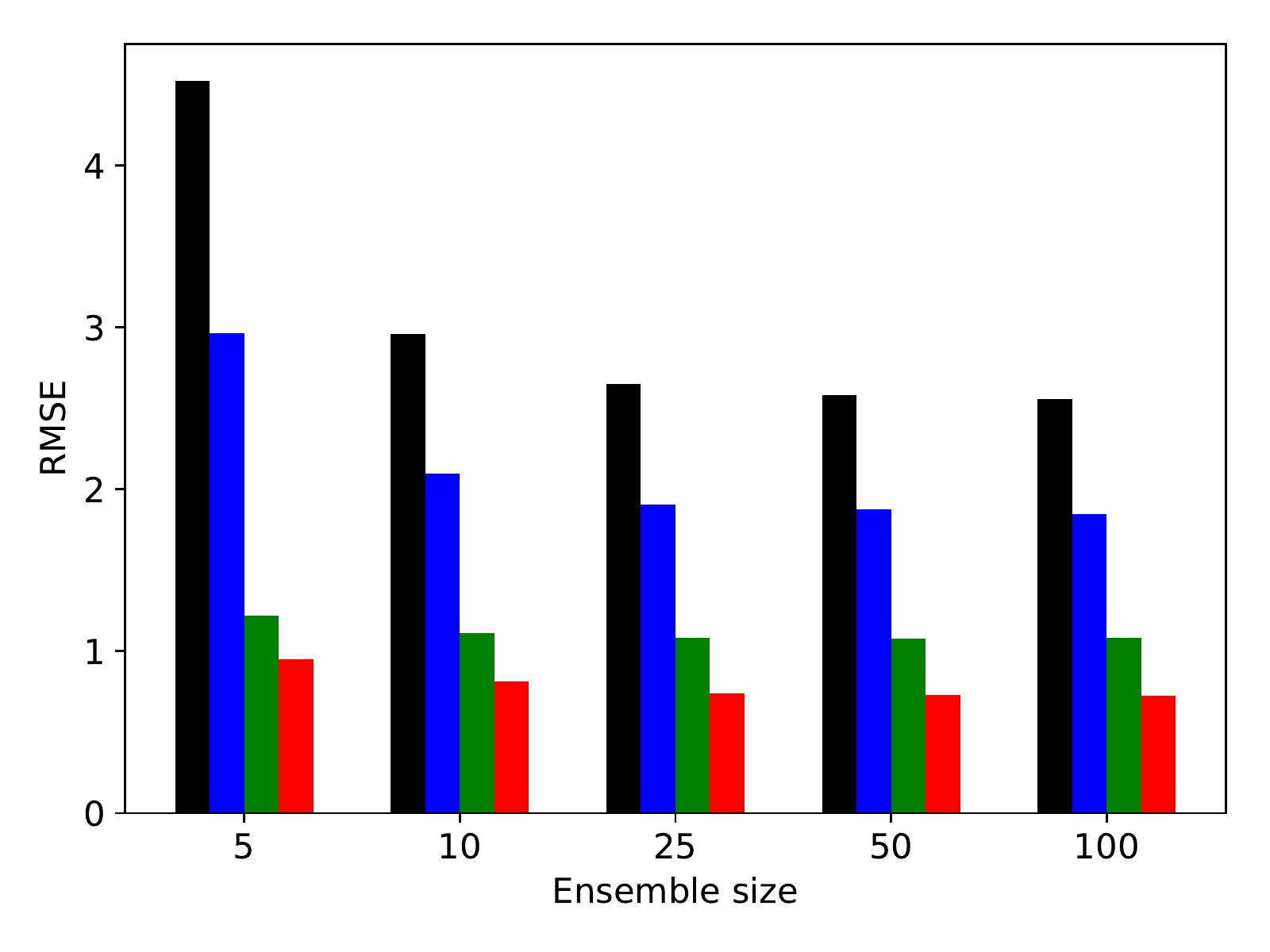} \\
    (a) & (b)
    \end{tabular}
  \caption{Mean RMSE for the optimal inflation coefficient and localization radius for (a) the LR ensemble and (b) the ULR ensemble. The black bars stand for the EnKF-LR, the blue bars for the srda-cubic, the green bars for the srda-NN, and the red bars for the EnKF-HR.}
  \label{fig:histogram_rmse}
\end{figure}

Figure~\ref{fig:histogram_rmse} displays the mean RMSE of the different assimilation schemes for (a) the LR ensemble and (b) the ULR ensemble. For both ensembles, the EnKF-LR/ULR scheme (black bars) displays the worst results because of the model error and the poor quality of the observations. The SRDA-cubic scheme (blue bars) displays better results in terms of RMSE, compared to those of the EnKF-LR/ULR with a reduction of the RMSE of approximately 18\% for the LR ensemble and up to 34\% for the ULR ensemble with 5 members. The only difference between the EnKF-LR/ULR and the SRDA-cubic standing in the assimilation step (that of the SRDA-cubic is performed in the HR space), it shows the benefits retrieved from the assimilation of HR observations with a lower error than that of the LR/ULR observations. In particular, in the case of the LR ensemble of size 5, the use of HR observations allows the algorithm to converge. On the other hand, in the case of the ULR ensemble of size 5, the EnKF-LR still converges but the error is very large. The SRDA-NN scheme (green bars) displays better results than the EnKF-LR and the SRDA-cubic. The neural network downscaling, as it reduces the forecast error of the LR/ULR model, provides HR background states that are close to the realistic HR QG fields than those computed with the SRDA-cubic, see section \ref{sect:results}-\ref{sect:super-resolution}, which results in a better estimation of the true state. The reduction of RMSE compared to the EnKF-LR/ULR is about 38-40\% for the LR ensemble and ranges from 57\% from to 73\% for the ULR ensemble. In the  particular case of the LR ensemble of size 5, the SRDA-NN converges while the EnKF-LR does not and the mean RMSE is close to that of the EnKF-HR. Despite an important reduction of the LR/ULR model error, the emulated HR fields computed by the neural network are still filled with error, which results ultimately in worse results of the SRDA-NN compared to the EnKF-HR (red bars) but with a significantly smaller computational cost, see table \ref{tab:computational_cost}. The relative degradation of the RMSE is between 11 and 14\% for the LR ensemble, it is between 28 and 49\% for the ULR ensemble. For both the LR and the ULR ensembles, thanks to the choice of the optimal inflation coefficients and localization radii, the schemes converge for small ensemble, $N=5,10$, and increasing the ensemble size does not allow for a significant improvement of the results. 

\begin{figure}[ht]
\hspace{-0.5cm}
    \begin{tabular}{cc}
    \includegraphics[width=0.5\textwidth]{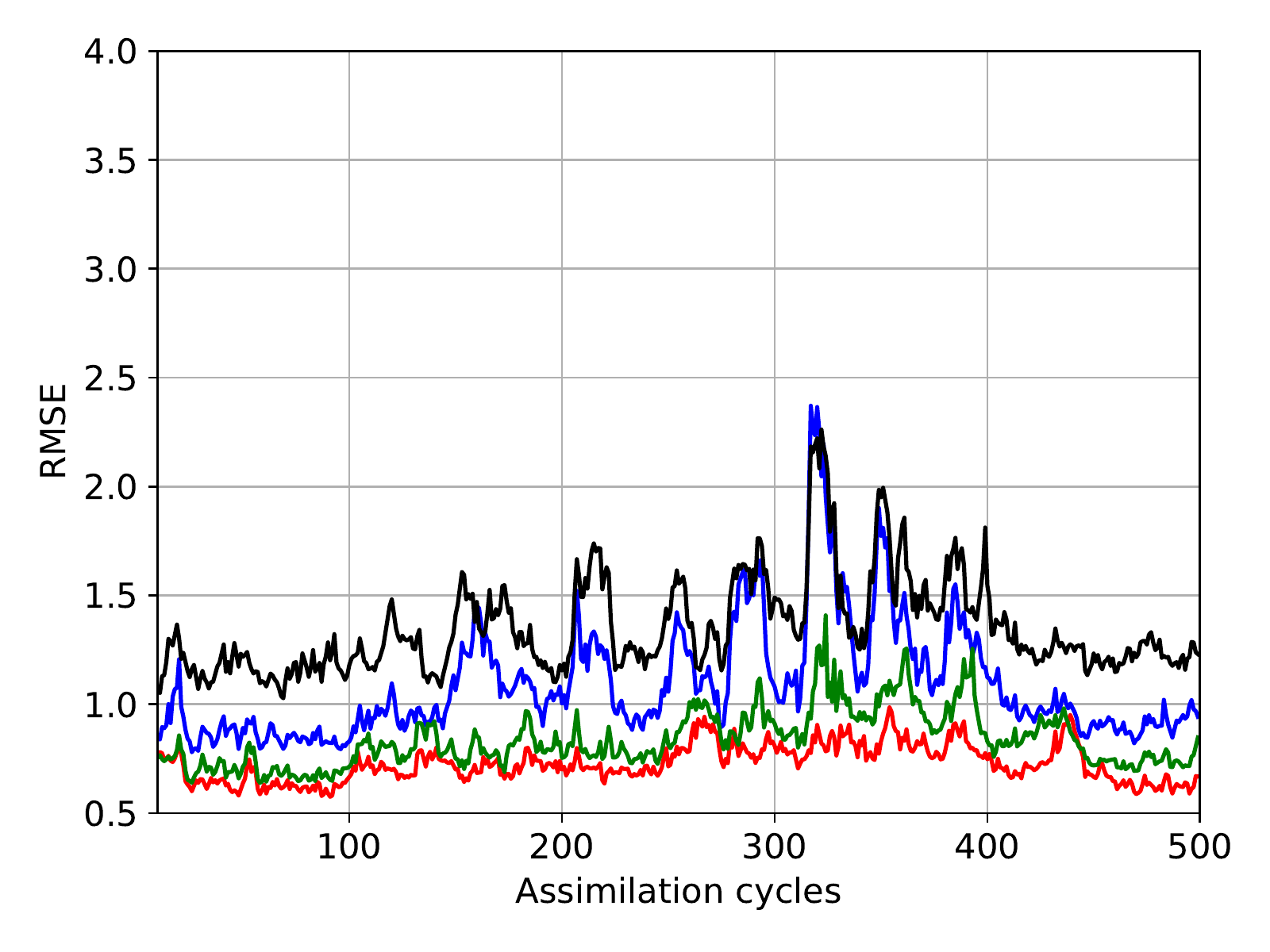} & \includegraphics[width=0.5\textwidth]{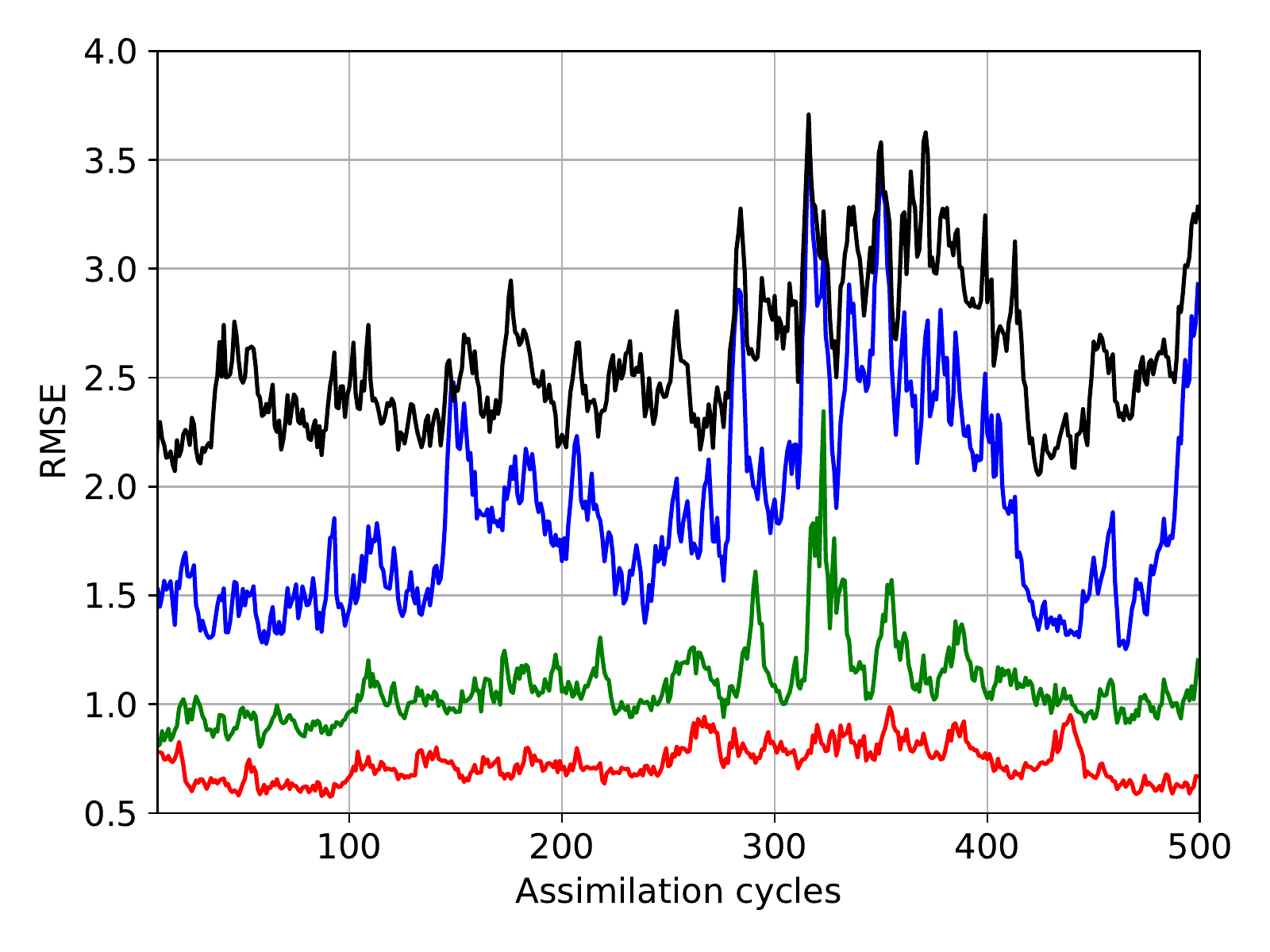} \\
    (a) & (b)
    \end{tabular}
  \caption{Time series of the rmse for (a) the LR ensemble and (b) the ULR ensemble and ensemble size $N=50$. The black lines stand for the EnKF-LR/ULR, the blue lines for the srda-cubic, the green lines for the srda-NN, and the red lines for the EnKF-HR.}
  \label{fig:evol_rmse}
\end{figure}

Figure~\ref{fig:evol_rmse} displays the time series of the RMSE for the different assimilation scheme and for ensemble size $N=50$. Figure~\ref{fig:evol_rmse}-(a) displays the results for the LR ensemble, Figure~\ref{fig:evol_rmse}-(b) displays the results for the ULR ensemble. At each assimilation cycle the relative performance of the schemes are the same as for figure figure~\ref{fig:histogram_rmse}. A similar conclusion stands for other ensemble sizes (results not showed). The EnKF-LR/ULR performs worse than all the other schemes. The SRDA-cubic performs worse than the SRDA-NN while the EnKF-HR performs better than all the other schemes. The RMSE of the SRDA-NN scheme displays a much smaller temporal variability, in particular during challenging events (for example around assimilation cycles 150 or between assimilation cycles 300-350), than the EnKF-LR/ULR and the SRDA-cubic schemes. This demonstrates the ability of the NN-scheme to cope with challenging events compared to the the two aforementioned schemes. In particular, while the average improvement of the mean RMSE of the srda-NN scheme was approximately of 37\% compared to the EnKF-LR and 24\% compared to the SRDA-cubic with the LR ensemble, the reduction of the RMSE of the SRDA-NN compared to the two previous schemes can reach almost 45\% during challenging events like the one between assimilation cycles 300-350 (figure~\ref{fig:evol_rmse}-(a)). 

\begin{figure}[ht]
\hspace{-0.5cm}
    \begin{tabular}{cc}
    \includegraphics[width=0.5\textwidth]{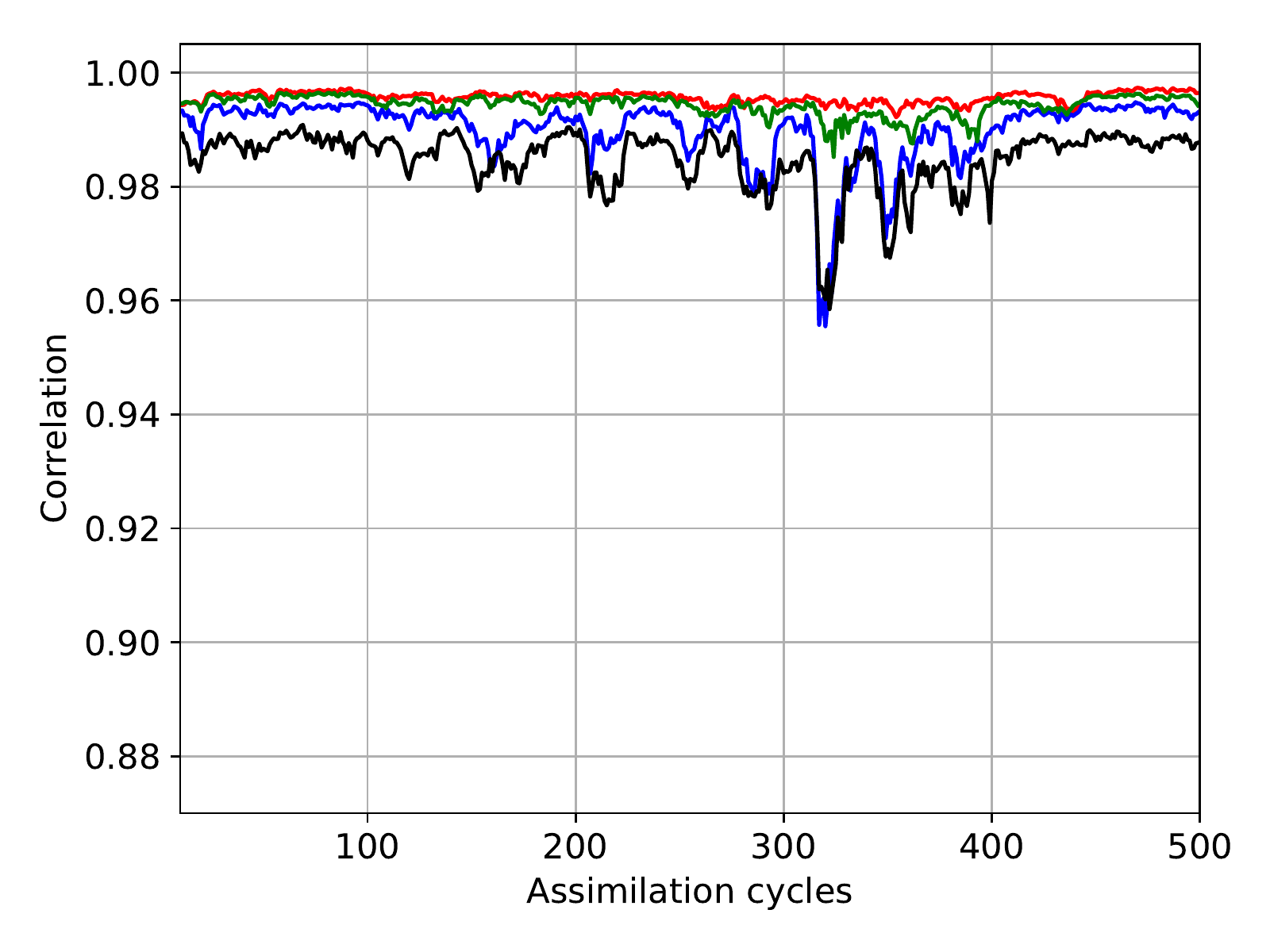} & \includegraphics[width=0.5\textwidth]{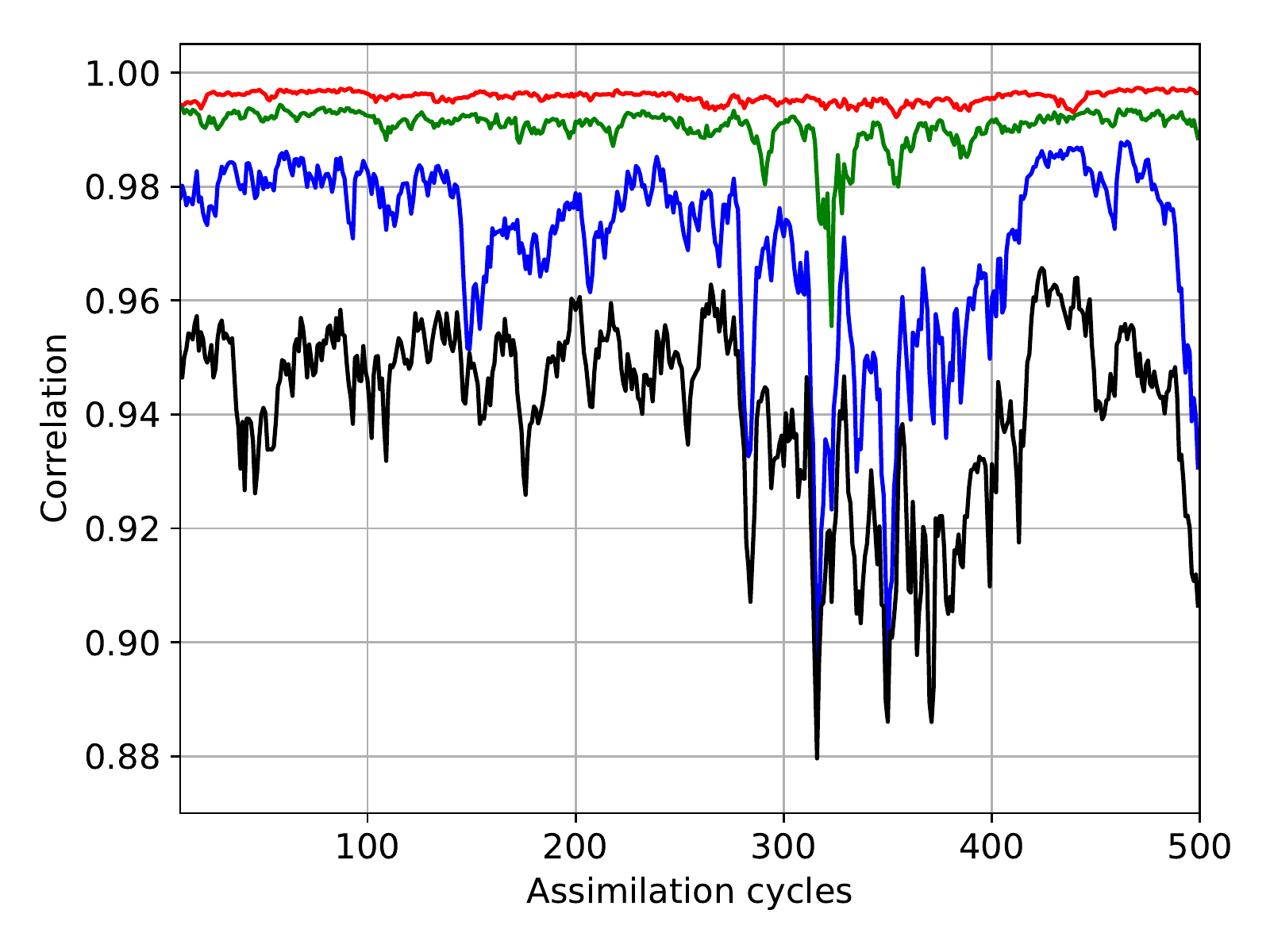} \\
    (a) & (b)
    \end{tabular}
    \caption{Time series of the Pearson correlation coefficient for (a) the LR ensemble and (b) the ULR ensemble and ensemble size $N=50$. The black line stands for the EnKF-LR/ULR, the blue lines for the srda-cubic, the green lines for the srda-NN, and the red lines for the EnKF-HR.}
  \label{fig:evol_correlation}
\end{figure}

Figure~\ref{fig:evol_correlation} displays the time series of the Pearson correlation coefficient for ensemble size $N=50$. The conclusions of the relative performances of the schemes remain the same as for the time series of the RMSE. The EnKF-LR/ULR and the EnKF-HR perform respectively the worst and the best while the SRDA-NN performs better than the SRDA-cubic. But the relative improvement of the correlation is not the same as before, as the correlation of all the schemes are larger than 0.95 for the LR ensemble and 0.88 for the ULR ensemble, meaning that all the schemes perform well in terms of correlation. It must be emphasized that the temporal evolution of the correlation follows a pattern that is anti-correlated to that of the RMSE, and that the improved performance of the SRDA-NN over the SRDA-cubic are in good agreement between correlation and RMSE. For example, between assimilation cycles 300-350 for the LR ensemble, there is a drop of the correlation for the EnKF-LR and the SRDA-cubic schemes while for the same period there is an increase of RMSE. 

\begin{figure}[ht]
\hspace{-0.5cm}
    \begin{tabular}{cc}
    \includegraphics[width=0.5\textwidth]{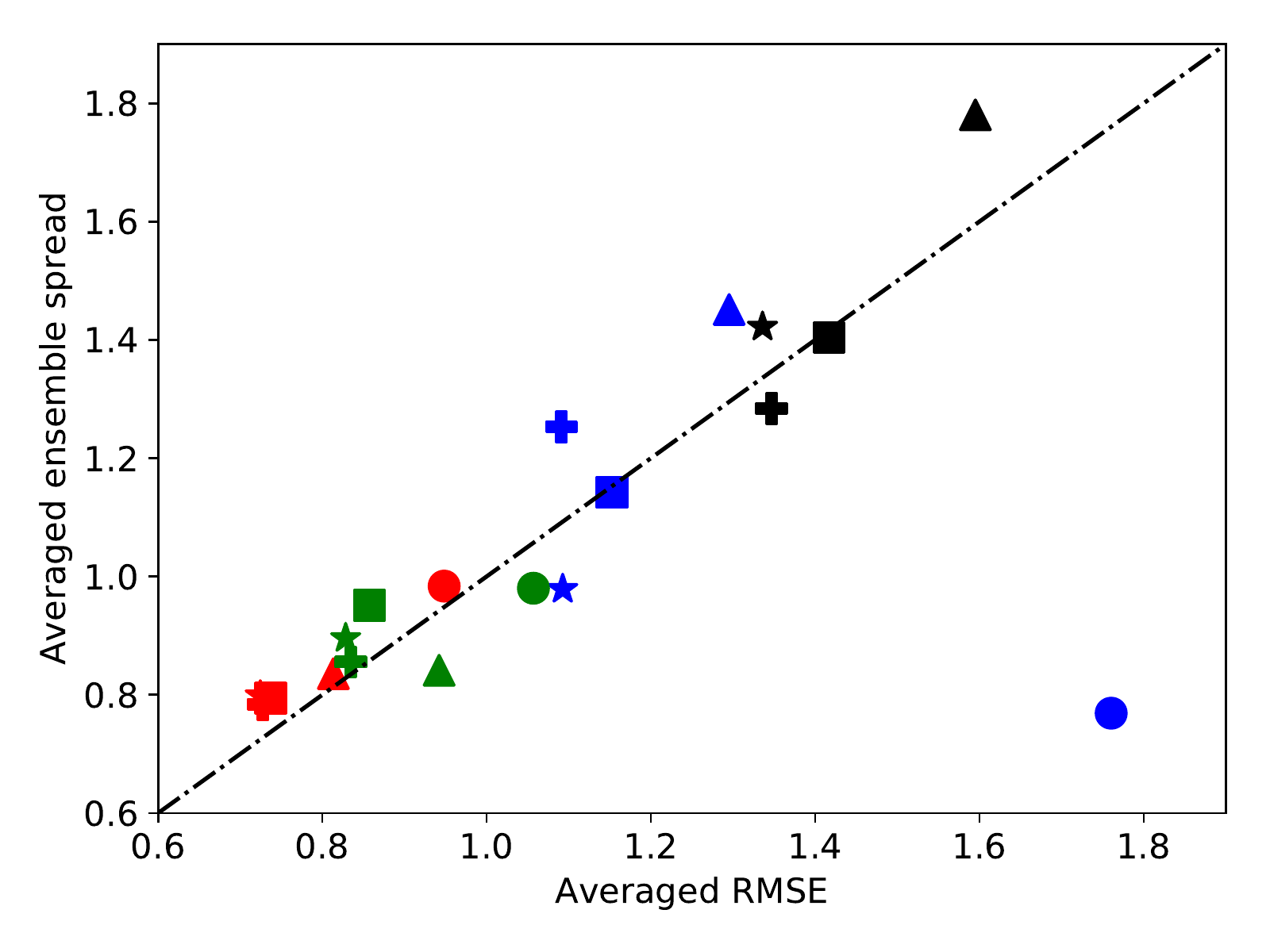} & \includegraphics[width=0.5\textwidth]{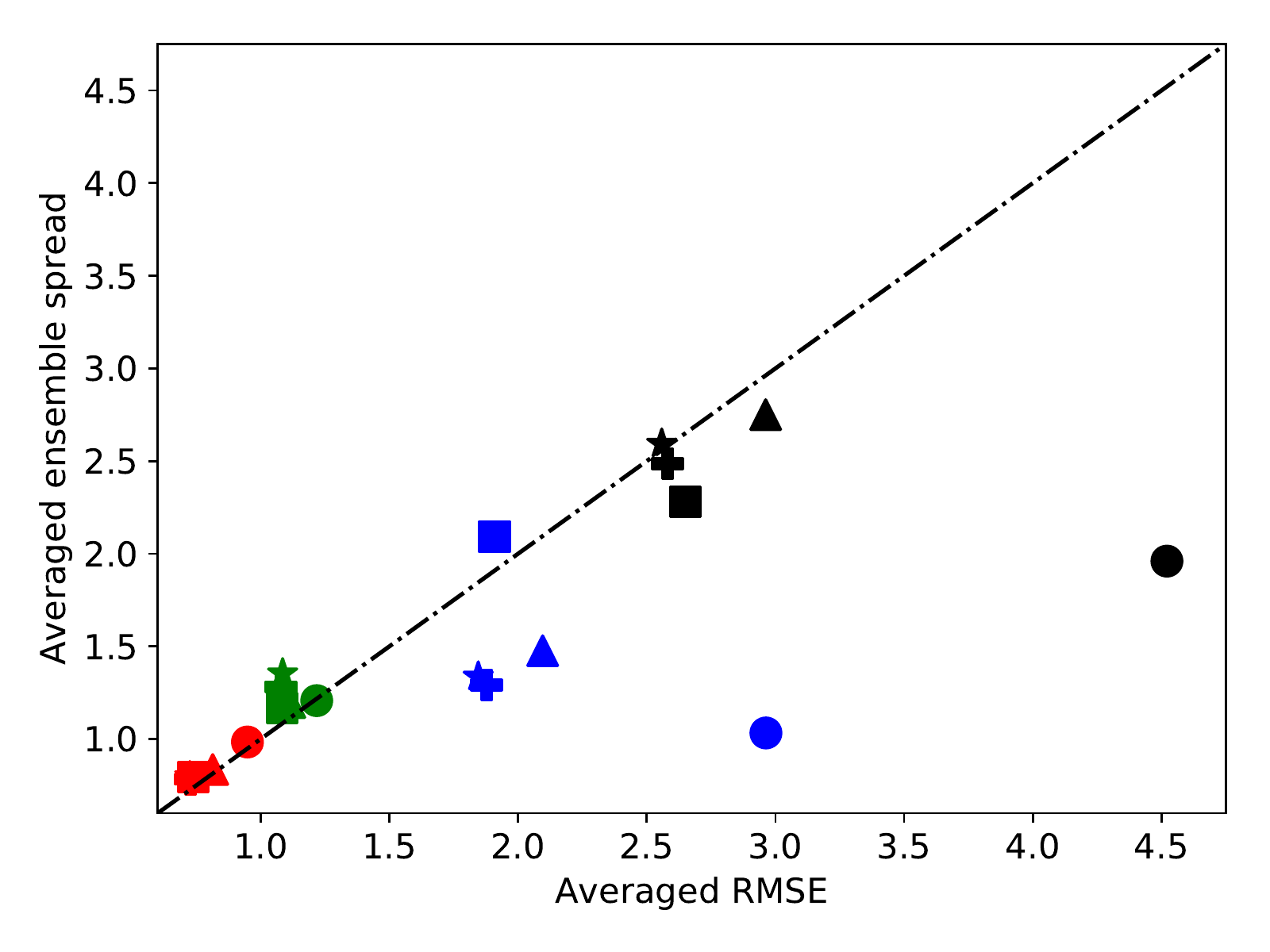} \\
    (a) & (b)
    \end{tabular}
    \caption{Averaged RMSE versus averaged ensemble spread for (a) the LR ensemble and (b) the ULR ensemble. The black symbols stand for the EnKF-LR, the blue ones for the srda-cubic, the green ones for the srda-NN, and the red ones for the EnKF-HR. The circles stand for ensemble size $N=5$, the triangles for $N=10$, the squares for $N=25$, the crosses for $N=50$ and the stars for $N=100$.}
  \label{fig:rmse_vs_spread}
\end{figure}

Figure~\ref{fig:rmse_vs_spread} displays the averaged RMSE versus the averaged ensemble spread for each ensemble size and each assimilation scheme, for the LR ensemble, figure~\ref{fig:rmse_vs_spread}-(a), and for the ULR ensemble, figure~\ref{fig:rmse_vs_spread}-(b). For each configuration (ensemble size, assimilation scheme), a point whose coordinates are the averaged RMSE and the averaged ensemble spread is plotted. In an ensemble prediction system, the ensemble spread should match the error of the ensemble mean \citep{Fortin2014,Rodwell2016}, if a configuration displays a correct relation spread/error, the corresponding point should be close to the diagonal (black dashed line). For the optimal parameters, all the schemes display a good spread/error relation for the different ensemble sizes, except for the case of $N=5$ for the SRDA-cubic and the EnKF-LR/ULR where the spread of the ensemble is under-estimated (in the sense that it is smaller than the RMSE). This implies in particular that the SRDA-NN preserves the reliability of the system whatever the size of the ensemble. More generally, if we consider the results of the whole sensitivity analysis (results not showed), the SRDA-NN  displays a spread close to that of the EnKF-HR, and that is much smaller than that of the SRDA-cubic and the EnKF-LR/ULR.

\begin{figure}[ht]
\hspace{-0.5cm}
    \begin{tabular}{c}
    \includegraphics[width=\textwidth]{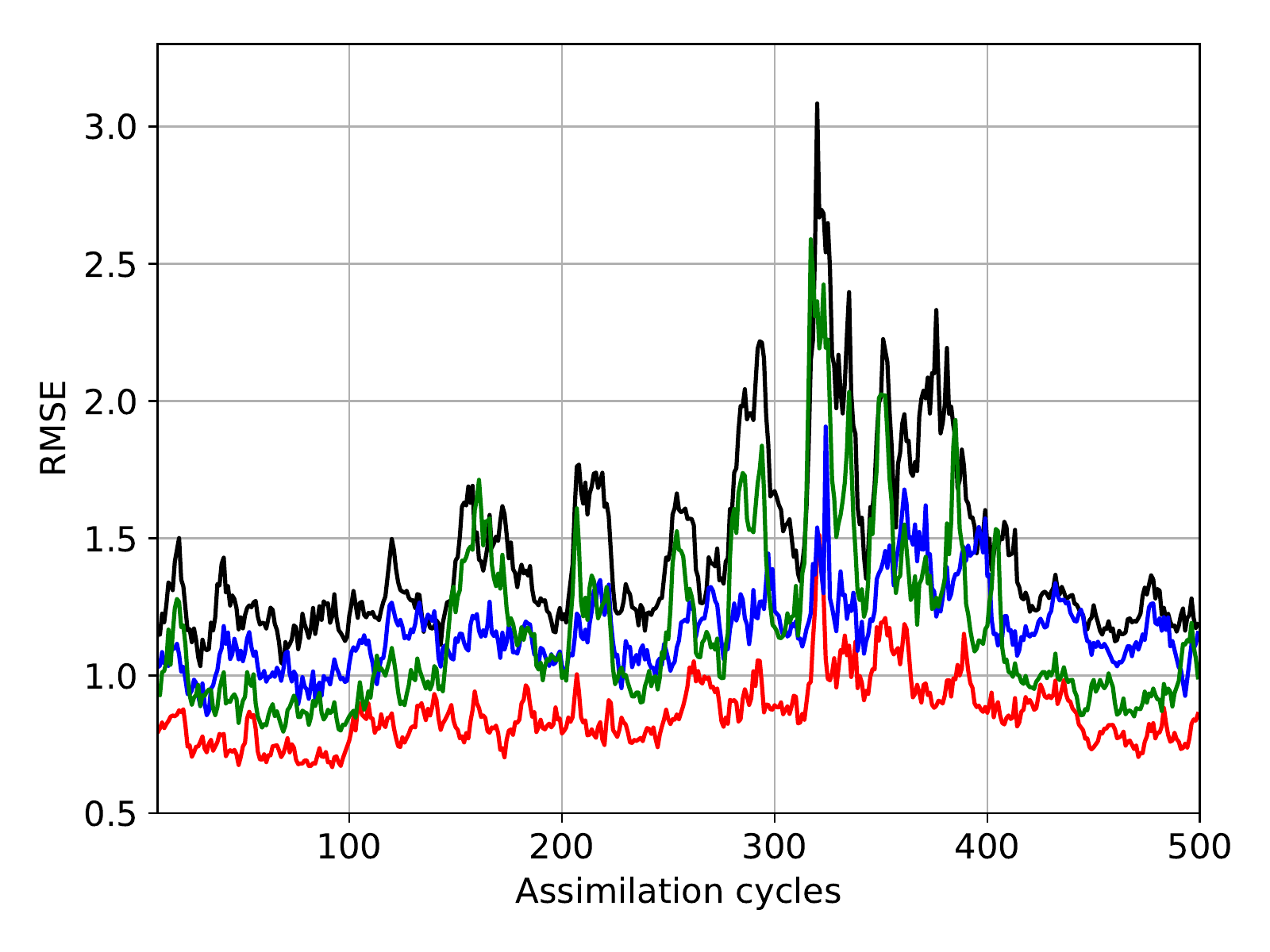}
    \end{tabular}
    \caption{Time series of the RMSE for the EnKF-LR (black line), the SRDA (red line), Eq.~(\ref{eq:srda_lr_mean})-(\ref{eq:srda_lr_anom}) with only model correction (blue line) and with only the super-resolution observation operator (green line).}
  \label{fig:model_correction_vs_super_resolution}
\end{figure}

The set of Eq.~(\ref{eq:srda_lr_mean})-(\ref{eq:srda_lr_anom}) was also tested and compared to the SRDA upscaled to the LR grid. Results (not presented here) show that the SRDA and this set equations provided similar results in terms of analyzed state. Rewriting the SRDA as a LR scheme with Eq.~(\ref{eq:srda_lr_mean})-(\ref{eq:srda_lr_anom}) allows to estimate and compare the respective impacts of the LR model error correction and the the super-resolution observation operator. Figure~\ref{fig:model_correction_vs_super_resolution} displays the time series of the RMSE for an ensemble of size 25, for the EnKF-LR (black line), the SRDA (red line), Eq.~(\ref{eq:srda_lr_mean})-(\ref{eq:srda_lr_anom}) with only model error correction (blue line), and Eq.~(\ref{eq:srda_lr_mean})-(\ref{eq:srda_lr_anom}) with only the super-resolution observation operator (gree line). Figure~\ref{fig:model_correction_vs_super_resolution} shows that both this configurations perform better than the EnKF-LR and worse than the SRDA. The SRDA providing similar results as the combination of correction of the model error and the super-resolution observation operator, figure~\ref{fig:model_correction_vs_super_resolution} shows that the combination of the two is necessary to achieve good results. In particular, we note that the correction of the model error achieves more stable results compared to the super-resolution observation operator and keeps the error small during challenging events (for example between assimilation cycles 300 and 350). On the other hand, the super-resolution observation operator achieves better performance outside of challenging events like for example before assimilation cycle 150, or after assimilation cycle 400.

\section{Discussion}\label{sect:discussion}

In this section, we discuss the potentiality of the method to scale up to more realistic models focusing in the foreseen challenges.

The method aims at being apply to high-dimension systems as it provides accurate high-resolution analysis at the cost of a low-dimension model. The costs of the algorithms are summarized in table~\ref{tab:cost}. All the experiments were done on an Intel Broadwell chips CPU with 64 GiB of memory except the training of the neural net that was computed on a NVIDIA V100 GPUs.
The small overhead of SRDA-NN compared with the cost of SRDA-cubic is due to the NN prediction which is more costly that the cubic interpolation.

\begin{table}[ht]
\caption{Cost of the different algorithm}\label{tab:cost}
\begin{center}
\begin{tabular}{ccc}
\hline\noalign{\smallskip}
Algorithm name & Number of model integrations & Wallclock time\\
EnKF-HR & I & 27m20s. \\
EnKF-LR & I/8 & 6m03s. \\
SRDA-cubic (LR) & I/8 & 9m29s.  \\
SRDA-NN (LR) & I/8 & 9m32s. \\
EnKF-ULR & I/64 &  2m26s. \\
SRDA-cubic (ULR) & I/64 & 7m20s. \\
SRDA-NN (ULR) & I/64 & 7m37s. \\
\noalign{\smallskip}\hline\noalign{\smallskip}
HR simulation for training & 120,000 & \\
$\mathcal{D}_{HL}$ training (LR) & 0 & 8min14s \\
$\mathcal{D}_{HL}$ training (ULR) & 0 & 8min51s \\

\hline
\end{tabular}
\label{tab:computational_cost}
\end{center}
\end{table}

Note that the overall cost of SRDA is highly reduced compared to that of the EnKF-HR because of the limited time of integration of the LR ensemble. But there is a need of computing resources ahead of the assimilation to produce a high-resolution simulation and to train the neural network. This aspect should not be overlooked, but the cost of simulating and training the neural network is predictable, so it can be taken into account when designing the data assimilation system.

The results presented here are obtained in a univariate case (the variable in the QG model represents either a stream function or sea-surface elevation, \citep{Sakov2008}). Using this method for more realistic setups would need to adapt the downscaling operator for a multivariate state. In principle, this is straightforward either for the DA method, which was proved to be working for multivariate state vectors, or for the neural network that was designed for 3 output variables (R,G,B). In practice, some attention has to be put on fulfilling some physical balances of multivariate models. For instance, even small inconsistencies between salinity and temperature could induce some instabilities of the density profile. As To address this problem, there might be a need to introduce physical constraints in the neural network training~\cite[see, e.g.,][]{beucler2021enforcing}. Note that physical considerations can also arise when considering the predictors specified as an input of the super-resolution operator. For example, it is expected that non-flat bathymetry can impact the downscaling procedure, and it might have to be specify as an input of the NN.

Finally, in this work, we have considered to downscale the whole domain. However, the SRDA algorithm could be adapted to perform high-resolution analysis only in  nested sub-regions of the domain where the error is expected to be high, which correspond in our case to regions of high mesoscale activity (see Fig~\ref{fig:compare_rmse}). This potential new version of SRDA is enabled by the fact that the NN acts only locally on the domain, due to the use of convolutional layers~\citep{bocquet2019data}.

\section{Conclusion}\label{sect:conclusion}

In this study, we presented a new data assimilation scheme derived from the Ensemble Kalman Filter that embeds a super-resolution neural network into a data assimilation system. Here we have used the EnKF but I can be applied to other sequential data assimilation methods. This new scheme aims at benefiting from the different resolutions of a model: the forecast step is performed with the low-resolution version of the model because of its limited computational cost, while the analysis step is performed in the high-resolution space to benefiting optimally from the high resolution observations. The scheme is called "super-resolution data assimilation" (SRDA) because the resolution of the background ensemble is increased to perform the assimilation step in the high-resolution space. Two different downscaling operators were studied and compared: a straightforward cubic spline interpolation operator and a more fancy convolutional neural network operator that is usually used for the super-resolution of photographic images. The method was applied with low and high-resolution versions of the model, but also with ultra low and high-resolution of the model.

The neural network was trained with a data set of matching pairs between (ultra) low and high-resolution states and showed better performance to reconstruct high-resolution fields compared to the cubic spline interpolation operator. In particular, the neural network was able to correct the position of the eddies of the reconstructed fields, reducing the impact of the low-resolution model error. The downscaling performance were also better with the low-resolution compared to the ultra-low resolution. 

The downscaling performance showed to have an important impact on the data assimilation performance. Both SRDA with cubic spline interpolation and with neural network perform better than the standalone low-resolution EnKF. But the SRDA neural network, as it benefits from the model error correction, performs better than its counterpart, and provides results close to those of the standalone high-resolution EnKF, but with a much smaller computational cost.

Nonetheless, this study was carried out in an idealised framework, with a model with only one variable and only one layer. The limitations of this method and its applicability to a more realistic model have been exposed and some tracks to overcome those limitations have been stated.

\begin{acknowledgements}
The authors wish to thank Mao-Lin Shen (UiB) for the fruitful discussions. 
Julien Brajard and François Counillon are funded by the project SFE(\#2700733 and \#309562) of the Norwegian Research Council. Julien Brajard is also associate professor at Sorbonne Université. Sébastien Barthélémy received support from EU H2020 Blue-Action (727852), the Trond Mohn Foundation, under project number BFS2018TMTO1. This work has also received a grant for computer time from the Norwegian Program for supercomputing (NOTUR2, project number nn9039k) and a storage grant (NORSTORE, NS9039k). 
\end{acknowledgements}

\appendix
\section{Rewriting the SRDA as a LR scheme}\label{app:lr_srda}
\label{app:}
If the LR and the HR grids overlap, the upscaling operator $\mathcal{U}$ is equivalent to a sub-sampling operator and is then linear. Applying $\mathcal{U}$ to the system of Eq.~(\ref{eq:update_mean_HR})-(\ref{eq:update_anomalies_HR}) leads to:

\begin{subequations}
\begin{equation}
    \mathbf{x}_{L,k}^{\rm a} = \mathcal{U}\left( \mathbf{x}_{H,k}^{\rm f}\right) + \frac{\mathcal{U}\left( \mathbf{A}_{H,k}^{\rm f}\right) \left( \mathbf{H}_{H,k} \mathbf{A}_{H,k}^{\rm f}\right)^T}{N-1} \left( \frac{\mathbf{H}_{H,k} \mathbf{A}_{H,k}^{\rm f} \left( \mathbf{H}_{H,k} \mathbf{A}_{H,k}^{\rm f}\right)^T}{N-1} + \mathbf{R} \right)^{-1} \left( \mathbf{d}_{H,k} - \mathbf{H}_{H,k} \mathbf{x}_{H,k}^{\rm f}\right), \label{eq:U_update_mean}
\end{equation}
\begin{equation}
    \mathbf{A}_{L,k}^{\rm a} =\mathcal{U}\left( \mathbf{A}_{H,k}^{\rm f}\right) -\frac{1}{2} \frac{\mathcal{U}\left( \mathbf{A}_{H,k}^{\rm f}\right) \left( \mathbf{H}_{H,k} \mathbf{A}_{H,k}^{\rm f}\right)^T}{N-1} \left( \frac{\mathbf{H}_{H,k} \mathbf{A}_{H,k}^{\rm f} \left( \mathbf{H}_{H,k} \mathbf{A}_{H,k}^{\rm f}\right)^T}{N-1} + \mathbf{R} \right)^{-1} \mathbf{H}_{H,k} \mathbf{A}_{H,k}^{\rm f},\label{eq:U_update_anom}
\end{equation}
\end{subequations}

\parindent=0cm where $\mathbf{x}_{L,k}^{\rm a} = \mathcal{U}\left( \mathbf{x}_{H,k}^{\rm a}\right)$ and $ \mathbf{A}_{L,k}^{\rm a} = \mathcal{U}\left( \mathbf{A}_{H,k}^{\rm a}\right)$.

\parindent=0.3cm $\mathbf{x}_{H,k}^{\rm f}$ and $\mathbf{A}_{H,k}^{\rm f}$ being respectively the mean and the anomalies of the downscaled background ensemble $\mathbf{E}_{H,k}^{\rm f} = \mathcal{D}\left( \mathbf{E}_{L,k}^{\rm f}\right)$ and $\mathcal{U}$ and $\mathcal{H}$ being linear, we have:

\begin{align*}
    \mathcal{U}\left( \mathbf{x}_{H,k}^{\rm f}\right) &= \overline{\mathcal{U}\circ \mathcal{D} \left( \mathbf{E}_{L,k}^{\rm f} \right)}, \\
    \mathcal{U}\left( \mathbf{A}_{H,k}^{\rm f}\right) &= \mathcal{U}\circ \mathcal{D} \left( \mathbf{E}_{L,k}^{\rm f}\right)-\overline{\mathcal{U}\circ \mathcal{D} \left( \mathbf{E}_{L,k}^{\rm f} \right)}, \\
    \mathbf{H}_{H,k} \mathbf{x}_{H,k}^{\rm f} &= \overline{\mathbf{H}_{H,k}\circ \mathcal{D} \left( \mathbf{E}_{L,k}\right)}, \\
    \mathbf{H}_{H,k} \mathbf{A}_{H,k}^{\rm f} &= \mathbf{H}_{H,k}\circ \mathcal{D} \left( \mathbf{E}_{L,k}\right) - \overline{\mathbf{H}_{H,k}\circ \mathcal{D} \left( \mathbf{E}_{L,k}\right)}
\end{align*}

For the sake of simplicity, in the following we define the operators $\mathcal{Q}$ and $\mathcal{H}$:

\begin{subequations}
\begin{equation}
\mathcal{Q}:\left\{
\begin{array}{ccc}
\mathbb{R}^{n_L\times N} & \rightarrow & \mathbb{R}^{n_L\times N} \\
\mathbf{E}_{L,k} & \mapsto & \tilde{\mathbf{E}}_{L,k} = \mathcal{U} \circ \mathcal{D}\left(\mathbf{E}_{L,k}\right)
\end{array}
\right.
\label{eq:definition_Q}
\end{equation}
\begin{equation}
\mathcal{H}:\left\{
\begin{array}{ccc}
\mathbb{R}^{n_L\times N} & \rightarrow & \mathbb{R}^{p\times N} \\
\mathbf{E}_{L,k} & \mapsto & \hat{\mathbf{E}}_{H,k} = \mathbf{H}_{H,k} \circ \mathcal{D}\left(\mathbf{E}_{L,k}\right)
\end{array}
\right.
\label{eq:definition_H}
\end{equation}
\end{subequations}

$\mathcal{Q}$ is an operator that maps the LR background ensemble $\mathbf{E}_{L,k}$ into the LR space and can be interpreted as an operator that corrects the LR model error, see section~\ref{sect:super-resolution}. In Eq.~(\ref{eq:definition_Q}), $\mathcal{Q}$ is defined as the composition of the operators $\mathcal{U}$ and $\mathcal{D}$, but it could be defined as an operator that minimizes the mean absolute error between the output and the training set upscaled to the LR grid. $\mathcal{U}\left( \mathbf{x}_{H,k}^{\rm f}\right)$ and $\mathcal{U}\left( \mathbf{A}_{H,k}^{\rm f}\right)$ represent then the mean and the anomalies of the corrected ensemble $\tilde{\mathbf{E}}_{L,k}$. In the following we note:

\begin{align}
    \mathcal{U}\left( \mathbf{x}_{H,k}^{\rm f}\right) &= \overline{\tilde{\mathbf{E}}_{L,k}} = \tilde{\mathbf{x}}_{L,k}^{\rm f}, \label{eq:Ux}\\
    \mathcal{U}\left( \mathbf{A}_{H,k}^{\rm f}\right) &= \tilde{\mathbf{E}}_{L,k}-\overline{\tilde{\mathbf{E}}_{L,k}} = \tilde{\mathbf{A}}_{L,k}^{\rm f}. \label{eq:UA}
\end{align}

$\mathcal{H}$ is an operator that maps the LR background ensemble $\mathbf{E}_{L,k}$ into the HR observation space and can be interpreted as a super-resolution observation operator. In Eq.~(\ref{eq:definition_H}), $\mathcal{H}_{H,k}$ is defined using the operator $\mathcal{D}$, but it could also be defined by proceeding to the super-resolution of the LR ensemble $\mathbf{E}_{L,k}$ at the observation points. $\mathbf{H}_{H,k} \mathbf{x}_{H,k}^{\rm f}$ and $\mathbf{H}_{H,k} \mathbf{A}_{H,k}^{\rm f}$ represent respectively the mean and the anomalies, at the HR observation points, of the downscaled LR ensemble $\mathbf{E}_{L,k}$. In the following, we note:

\begin{align}
    \mathbf{H}_{H,k} \mathbf{x}_{H,k}^{\rm f} &= \overline{\hat{\mathbf{E}}_{H,k}^{\rm f}} = \hat{\mathbf{x}}_{H,k}^{\rm f}, \label{eq:Hx}\\
    \mathbf{H}_{H,k} \mathbf{A}_{H,k}^{\rm f} &= \hat{\mathbf{E}}_{H,k}^{\rm f}-\overline{\hat{\mathbf{E}}_{H,k}^{\rm f}} = \hat{\mathbf{A}}_{H,k}^{\rm f}.\label{eq:HA}
\end{align}

Replacing the terms $\mathcal{U}\left( \mathbf{x}_{H,k}^{\rm f}\right)$, $\mathcal{U}\left( \mathbf{A}_{H,k}^{\rm f}\right)$, $\mathbf{H}_{H,k} \mathbf{x}_{H,k}^{\rm f}$, and $\mathbf{H}_{H,k} \mathbf{A}_{H,k}^{\rm f}$ in Eq.~(\ref{eq:U_update_mean})-(\ref{eq:U_update_mean}), respectively by $\tilde{\mathbf{x}}_{L,k}^{\rm f}$, $\tilde{\mathbf{A}}_{L,k}^{\rm f}$, $\hat{\mathbf{x}}_{H,k}^{\rm f}$, and $\hat{\mathbf{A}}_{H,k}^{\rm f}$, we get Eq.~(\ref{eq:srda_lr_mean})-(\ref{eq:srda_lr_anom}).

\newpage


%


%
%

\bibliographystyle{spbasic}      
\bibliography{MyCollection, references}   

\begin{thebibliography}{25}
\providecommand{\natexlab}[1]{#1}
\providecommand{\url}[1]{{#1}}
\providecommand{\urlprefix}{URL }
\expandafter\ifx\csname urlstyle\endcsname\relax
  \providecommand{\doi}[1]{DOI~\discretionary{}{}{}#1}\else
  \providecommand{\doi}{DOI~\discretionary{}{}{}\begingroup
  \urlstyle{rm}\Url}\fi
\providecommand{\eprint}[2][]{\url{#2}}

\bibitem[{Beucler et~al.(2021)Beucler, Pritchard, Rasp, Ott, Baldi, and
  Gentine}]{beucler2021enforcing}
Beucler T, Pritchard M, Rasp S, Ott J, Baldi P, Gentine P (2021) Enforcing
  analytic constraints in neural networks emulating physical systems. Physical
  Review Letters 126(9):098302

\bibitem[{Bocquet et~al.(2019)Bocquet, Brajard, Carrassi, and
  Bertino}]{bocquet2019data}
Bocquet M, Brajard J, Carrassi A, Bertino L (2019) Data assimilation as a
  learning tool to infer ordinary differential equation representations of
  dynamical models. Nonlinear Processes in Geophysics 26(3):143--162

\bibitem[{Brajard et~al.(2021)Brajard, Carrassi, Bocquet, and
  Bertino}]{brajard2021combining}
Brajard J, Carrassi A, Bocquet M, Bertino L (2021) Combining data assimilation
  and machine learning to infer unresolved scale parametrization. Philosophical
  Transactions of the Royal Society A 379(2194):20200086

\bibitem[{Evensen(2003)}]{Evensen2003}
Evensen G (2003) {The Ensemble Kalman Filter: Theoretical formulation and
  practical implementation}. Ocean Dynamics 53(4):343--367,
  \doi{10.1007/s10236-003-0036-9}

\bibitem[{Fortin et~al.(2014)Fortin, Abaza, Anctil, and Turcotte}]{Fortin2014}
Fortin V, Abaza M, Anctil F, Turcotte R (2014) {Why Should Ensemble Spread
  Match the RMSE of the Ensemble Mean ?} Journal of Hydrometeorology
  15(2010):1708--1714, \doi{10.1175/JHM-D-14-0008.1}

\bibitem[{Gao and Xue(2008)}]{Gao2008}
Gao J, Xue M (2008) {An Efficient Dual-Resolution Approach for Ensemble Data
  Assimilation and Tests with Simulated Doppler Radar Data}. Monthly Weather
  Review 136(3):945--963, \doi{10.1175/2007MWR2120.1},
  \urlprefix\url{http://journals.ametsoc.org/doi/abs/10.1175/2007MWR2120.1}

\bibitem[{Gent et~al.(1995)Gent, Willebrand, McDougall, and
  McWilliams}]{Gent1995}
Gent PR, Willebrand J, McDougall TJ, McWilliams JC (1995) {Parameterizing
  Eddy-Induced Tracer Transports in Ocean Circulation Models}. Journal of
  physical oceanography 25(4):463--474

\bibitem[{Hallberg(2013)}]{Hallberg2013}
Hallberg R (2013) {Using a resolution function to regulate parameterizations of
  oceanic mesoscale eddy effects}. Ocean Modelling 72:92--103,
  \doi{10.1016/j.ocemod.2013.08.007},
  \urlprefix\url{http://dx.doi.org/10.1016/j.ocemod.2013.08.007}

\bibitem[{He et~al.(2016)He, Zhang, Ren, and Sun}]{he2016deep}
He K, Zhang X, Ren S, Sun J (2016) Deep residual learning for image
  recognition. In: Proceedings of the IEEE conference on computer vision and
  pattern recognition, pp 770--778

\bibitem[{Hewitt et~al.(2017)Hewitt, Bell, Chassignet, Czaja, Ferreira, Gri,
  Hyder, Mcclean, New, and Roberts}]{Hewitt2017}
Hewitt HT, Bell MJ, Chassignet EP, Czaja A, Ferreira D, Gri SM, Hyder P,
  Mcclean JL, New AL, Roberts MJ (2017) {Will high-resolution global ocean
  models bene fi t coupled predictions on short-range to climate timescales ?}
  120(July):120--136, \doi{10.1016/j.ocemod.2017.11.002}

\bibitem[{Hunt et~al.(2007)Hunt, Kostelich, and Szunyogh}]{Hunt2007}
Hunt BR, Kostelich EJ, Szunyogh I (2007) {Efficient data assimilation for
  spatiotemporal chaos: A local ensemble transform Kalman filter}. Physica D:
  Nonlinear Phenomena 230:112--126, \doi{10.1016/j.physd.2006.11.008},
  \eprint{0511236}

\bibitem[{Janji{\'{c}} et~al.(2018)Janji{\'{c}}, Bormann, Bocquet, Carton,
  Cohn, Dance, Losa, Nichols, Potthast, Waller, and Weston}]{Janjic2018}
Janji{\'{c}} T, Bormann N, Bocquet M, Carton JA, Cohn SE, Dance SL, Losa SN,
  Nichols NK, Potthast R, Waller JA, Weston P (2018) {On the representation
  error in data assimilation}. Quarterly Journal of the Royal Meteorological
  Society 144(713):1257--1278, \doi{10.1002/qj.3130}

\bibitem[{Kingma and Ba(2014)}]{kingma2014adam}
Kingma DP, Ba J (2014) Adam: A method for stochastic optimization. arXiv
  preprint arXiv:14126980

\bibitem[{LeCun et~al.(1989)}]{lecun1989generalization}
LeCun Y, et~al. (1989) Generalization and network design strategies.
  Connectionism in perspective 19:143--155

\bibitem[{Lei and Whitaker(2017)}]{lei2017}
Lei L, Whitaker JS (2017) {Journal of Advances in Modeling Earth Systems}.
  Journal of Advances in Modeling Earth Systems 9:781--789,
  \doi{10.1002/2017MS001065}

\bibitem[{Lim et~al.(2017)Lim, Son, Kim, Nah, and
  Mu~Lee}]{Lim_2017_CVPR_Workshops}
Lim B, Son S, Kim H, Nah S, Mu~Lee K (2017) Enhanced deep residual networks for
  single image super-resolution. In: Proceedings of the IEEE Conference on
  Computer Vision and Pattern Recognition (CVPR) Workshops

\bibitem[{Rainwater and Hunt(2013)}]{Rainwater2013}
Rainwater S, Hunt B (2013) {Mixed-Resolution Ensemble Data Assimilation}.
  Monthly Weather Review 141(9):3007--3021, \doi{10.1175/mwr-d-12-00234.1}

\bibitem[{Rodrigues et~al.(2018)Rodrigues, Oliveira, Cunha, and
  Netto}]{Rodrigues2018}
Rodrigues ER, Oliveira I, Cunha RLF, Netto MAS (2018) {DeepDownscale : a deep
  learning strategy for high-resolution weather forecast}
  \doi{10.1109/eScience.2018.00130}

\bibitem[{Rodwell et~al.(2016)Rodwell, Lang, Ingleby, Bormann, Holm, Rabier,
  Richardson, and Yamaguchi}]{Rodwell2016}
Rodwell MJ, Lang STK, Ingleby NB, Bormann N, Holm E, Rabier F, Richardson DS,
  Yamaguchi M (2016) {Reliability in ensemble data assimilation ´}. Quarterly
  Journal of the Royal Meteorological Society 142(January):443--454,
  \doi{10.1002/qj.2663}

\bibitem[{Sakov and Bertino(2011)}]{Sakov2011}
Sakov P, Bertino L (2011) {Relation between two common localisation methods for
  the EnKF}. Computational Geosciences 15(2):225--237,
  \doi{10.1007/s10596-010-9202-6}

\bibitem[{Sakov and Oke(2008)}]{Sakov2008}
Sakov P, Oke PR (2008) {A deterministic formulation of the ensemble Kalman
  filter: An alternative to ensemble square root filters}. Tellus, Series A:
  Dynamic Meteorology and Oceanography 60 A(2):361--371,
  \doi{10.1111/j.1600-0870.2007.00299.x}

\bibitem[{Sandery and Sakov(2017)}]{Sandery2017}
Sandery PA, Sakov P (2017) {Ocean forecasting of mesoscale features can
  deteriorate by increasing model resolution towards the submesoscale}. Nature
  Communications 8(1):1--8, \doi{10.1038/s41467-017-01595-0},
  \urlprefix\url{http://dx.doi.org/10.1038/s41467-017-01595-0}

\bibitem[{Shi et~al.(2016)Shi, Caballero, Husz{\'a}r, Totz, Aitken, Bishop,
  Rueckert, and Wang}]{shi2016real}
Shi W, Caballero J, Husz{\'a}r F, Totz J, Aitken AP, Bishop R, Rueckert D, Wang
  Z (2016) Real-time single image and video super-resolution using an efficient
  sub-pixel convolutional neural network. In: Proceedings of the IEEE
  conference on computer vision and pattern recognition, pp 1874--1883

\bibitem[{Thoppil et~al.(2021)Thoppil, Frolov, Rowley, Reynolds, Jacobs,
  Metzger, Hogan, Barton, Wallcraft, Smedstad, and Shriver}]{Thoppil2021}
Thoppil PG, Frolov S, Rowley CD, Reynolds CA, Jacobs GA, Metzger EJ, Hogan PJ,
  Barton N, Wallcraft AJ, Smedstad OM, Shriver JF (2021) prediction horizon for
  ocean mesoscale variability. Communications Earth \& Environment 2(1):1--9,
  \doi{10.1038/s43247-021-00151-5},
  \urlprefix\url{http://dx.doi.org/10.1038/s43247-021-00151-5}

\bibitem[{Vandal et~al.(2018)Vandal, Kodra, Ganguly, Michaelis, Nemani, and
  Ganguly}]{Vandal2018}
Vandal T, Kodra E, Ganguly S, Michaelis A, Nemani R, Ganguly AR (2018) {DeepSD
  : Generating High Resolution Climate Change Projections through Single Image
  Super-Resolution}. International Joint Conferences on Artificial Intelligence
  Organization pp 1663--1672

\end{thebibliography}

\end{document}